\documentclass[
  floats,
  floatfix,
  prd,
  twocolumn,
  reprint,
  superscriptaddress,
  nofootinbib,
  longbibliography
]{revtex4-2}
\usepackage{amsfonts,amssymb,amsmath}
\usepackage[caption=false]{subfig} %
\usepackage{graphicx, hyperref, tabularx, dcolumn}

\usepackage{color}
\usepackage[dvipsnames]{xcolor}
\usepackage[normalem]{ulem}
\usepackage{orcidlink}
\usepackage{tensor, comment}
\usepackage{enumitem}
\usepackage[export]{adjustbox}

\newcommand{\Caltech}{\affiliation{Theoretical Astrophysics 350-17, California
    Institute of Technology, Pasadena, CA 91125, USA}}
\newcommand{\CaltechBurke}{\affiliation{Walter Burke Institute for Theoretical Physics, California
    Institute of Technology, Pasadena, CA 91125, USA}}
\newcommand{\Cornell}{\affiliation{Cornell Center for Astrophysics and Planetary Science, Cornell University, Ithaca, NY 14853, USA}}
\newcommand{\CornellPhysics}{\affiliation{Department of Physics, Cornell University, Ithaca, NY, 14853, USA}}

\newcommand{\dual}{\,{}^*\!}
\newcommand{\spatiallc}[1]{\varepsilon_{(3)}^{#1}}
\newcommand{\e}[1]{\times 10^{#1}}
\newcommand{\dgpn}[1]{DG-$P_{#1}$}

\newcommand{\spectre}{\textsc{SpECTRE}}
\newcommand{\reffig}[1]{Fig.~\ref{#1}}
\newcommand{\refsec}[1]{Sec.~\ref{#1}}

\begin{document}

\author{Yoonsoo Kim \orcidlink{0000-0002-4305-6026}} \email{ykim7@caltech.edu} \Caltech 
\author{Elias R. Most \orcidlink{0000-0002-0491-1210}} \Caltech \CaltechBurke
\author{William Throwe \orcidlink{0000-0001-5059-4378}} \Cornell
\author{Saul A. Teukolsky \orcidlink{0000-0001-9765-4526}} \Caltech \CaltechBurke \Cornell
\author{Nils Deppe \orcidlink{0000-0003-4557-4115}} \CornellPhysics \Cornell

\title{General relativistic force-free electrodynamics with a discontinuous
    Galerkin-finite difference hybrid method}

\date{\today}

\begin{abstract}
  Relativistic plasmas around compact objects can sometimes be approximated as
  being force-free. In this limit, the plasma inertia is negligible and the
  overall dynamics is governed by global electric currents. We present a novel
  numerical approach for simulating such force-free plasmas, which allows for
  high accuracy in smooth regions as well as capturing dissipation in current
  sheets. Using  a high-order accurate discontinuous Galerkin method augmented
  with a conservative finite-difference method, we demonstrate efficient global
  simulations of black hole and neutron star magnetospheres. In addition to a
  series of challenging test problems, we show that our approach can---depending
  on the physical properties of the system and the numerical implementation---be
  up to $10\times$ more efficient than conventional simulations, with a speedup
  of 2-3$\times$ for most problems we consider in practice.
\end{abstract}

\maketitle

\section{Introduction}

Compact objects such as neutron stars and black holes can feature some of the
strongest magnetic fields in the universe. Under these conditions, the
environments surrounding them can be filled with a highly conducting plasma. The
plasma dynamics of these magnetospheres are thought to be responsible for
several observable transients in the radio
\cite{Bochenek:2020zxn,Lyubarsky2020,Mahlmann:2022nnz} and X-ray
\cite{Thompson:1995gw,Kaspi:2003xh,Beloborodov:2012ug,Archibald:2016npx,Tavani2021}
bands. While the description of emission processes fundamentally necessitates
modeling the relevant kinetic scales \cite{Philippov_2015,Chen_2017}, the
available energy budget as well as the presence of any dissipative or emitting
region inside the magnetosphere is a result of the bulk dynamics. It is this
latter aspect that our present work aims to advance.  Since these scenarios are
highly nonlinear, their effective description requires numerical approaches.

The global dynamics of the plasma is usually modeled under several simplifying
assumptions. In a very strongly magnetized magnetosphere, the inertia of the
plasma can approximately be neglected \cite{Uchida1997,Gruzinov1999}. In this
force-free electrodynamics (FFE) state, the evolution of the system is governed
largely by bulk currents, obtained via an effective closure of the Maxwell
equations. It is important to point out that the main assumption---neglecting
plasma inertia---can break down, e.g., during shock formation, as well as the
absence of physically meaningful dissipation in reconnection regions. In this
regime, the closest extension of force-free electrodynamics is
magnetohydrodynamical (MHD) models, retaining a single-component plasma
rest-mass density. MHD studies of relativistic magnetospheres are not commonly
employed (see, e.g., Ref. \cite{Tchekhovskoy_2013} for a notable exception in
neutron star magnetospheres). Instead, most studies adopt an FFE approach.

Recent examples include applications to magnetar quakes \cite{Bransgrove2020},
nonlinear steepening of Alfv\'en waves \cite{Yuan:2020ayo,Yuan:2022uqt},
magnetar giant flares
\cite{Parfrey:2013gza,Mahlmann:2019arj,Carrasco:2019aas,Mahlmann:2023ipm},
outbursts from gravitational collapse of a neutron star \cite{Lehner_2012} (see
also Refs. \cite{Nathanail:2017wly,Most:2018abt} for related studies in
electrovacuum), and black hole and neutron star magnetospheres
\cite{Komissarov2004,Spitkovsky2006,McKinney:2006sd,Chen:2020,Carrasco:2020sxg}.
Apart from isolated compact objects, force-free electrodynamics has also been
employed in the context of jets from massive black hole mergers
\cite{Palenzuela2010b,Palenzuela:2010nf} and potential electromagnetic precursor
to gravitational wave events involving merger of compact objects
\cite{Alic2012,Palenzuela2010a,Carrasco:2021,Most:2020ami,Most2022,Most2023,Most:2023unc}.

Several approaches have been adopted in the literature for numerically solving
the FFE equations. Most commonly, either unlimited finite-difference
\cite{Spitkovsky2006,Kalapotharakos:2008zc,Palenzuela2010b,Chen:2020,Carrasco:2017ucd}
or conservative finite-volume schemes
\cite{Komissarov2002,Cho:2004nn,Asano:2005di,McKinney2006,Yu2011,Etienne2017,Most:2020ami,Mahlmann:2021a}
have been employed. These methods are robust, can easily capture strong
gradients inside the magnetosphere, and work well with commonly employed mesh
refinement techniques \cite{2016arXiv161008833D}. However, they come with a
major drawback. Properly capturing wave solutions over long integration times,
e.g., Alfv\'en waves \cite{Yuan:2020ayo}, requires a large number of grid
points, especially when less accurate versions of finite-difference/volume
schemes are being used. This prohibitively increases computational costs,
especially for applications such as compact binary magnetospheres in which scale
separations can span two orders of magnitude.

On the other hand, spectral-type methods such as the pseudospectral method offer
exponential convergence for smooth solutions, providing a maximum in accuracy
over computational cost. Several studies have made use of spectral schemes to
solve the FFE equations \cite{Parfrey2012,Petri2012,Cao2015}. One serious
limitation of spectral methods is the appearance of unphysical oscillations
(Gibbs phenomenon) near a discontinuity or a large gradient e.g. current sheets,
which are naturally present in compact object magnetospheres. Remedying these
numerical instabilities requires special treatments such as filtering or a
limiting procedure \cite{hesthaven2007nodal}. In addition, globally spectral
methods are not easily parallelizable, making it difficult to simulate physical
scenarios with large scale separations.

To counteract this shortcoming, popular approaches in the literature focus on
spectral element methods in which the computational domain is divided into
non-overlapping spectral elements, communicating only with the directly
neighboring elements through the element boundaries. This approach allows for
highly parallelizable implementations while retaining the exponential
convergence property for smooth solutions. A concrete example of this approach,
a discontinuous Galerkin (DG) method, is gaining its popularity in computational
fluid dynamics and astrophysics
\cite[e.g.][]{Bugner:2015gqa,Zanotti2015,Zanotti:2015mia,Fambri:2018,Lombart2020,Reinarz_2020,Deppe2022,Tichy2023,Dumbser:2023see,2024arXiv240106841C},
as well as in FFE \cite{Petri:2015,Petri:2016}.

While a DG scheme naturally permits a discontinuity at the element boundary,
without special care to suppress unphysical oscillations, it suffers from the
same fate as globally spectral methods described above. Several strategies have
been proposed in the DG literature, which are frequently referred to as
\emph{limiters}. Common types of DG limiters are implemented as direct
manipulations on spectral coefficients, addition of artificial viscosity, or a
flattening correction of the solution with respect to its average value within
an element. We refer the reader to \cite{Zanotti2015,Deppe2021CQG} and
references therein for the available types of DG limiters and related
discussions.

DG limiters currently are not particularly accurate or reliable compared with
corresponding finite-volume or finite-difference techniques, especially for
curved meshes or relativistic applications (see, e.g.,
Ref.~\cite{Deppe2021CQG}). A recently developed alternative strategy is to
supplement the DG evolution with a more robust sub-element discretization, which
has been mostly chosen to be finite volume
\cite[e.g.][]{Dumbser2014,Zanotti2015,Zanotti:2015mia,Fambri:2018,
Vilar2019,Nunezdelarosa2018,RuedaRamirez2022,Maltsev2023}. Motivated by the idea
of the a posteriori finite-volume limiting approach of Ref. \cite{Dumbser2014},
the discontinuous Galerkin-finite difference (DG-FD) hybrid method was
introduced by \cite{Deppe2021CQG} (see also Ref. \cite{Deppe2022,Legred:2023zet}
for applications to relativistic fluid dynamics simulations).

In this paper, we present a new numerical scheme and code for
general-relativistic FFE simulations based on a discontinuous Galerkin
discretization. Our motivation is twofold. First, we explore the suitability of
the DG-FD hybrid approach to enable large-scale, parallel yet accurate numerical
simulations, especially of compact binary magnetospheres.  Second, since FFE on
physical grounds has very localized regions of non-smoothness such as current
sheets, these simulations serve as an ideal testbed to calibrate and assess the
usefulness of the DG-FD hybrid approach. Our hybrid scheme also incorporates
previously developed implicit-explicit time integration schemes
\cite{Pareschi2005} (see Refs. \cite{Most:2020ami,Most2022} for applications to
the FFE system), which allows us to enforce a set of algebraic constraints
present in the FFE system. This joint approach achieves high-order convergence
in smooth regions while capturing discontinuous features such as magnetic
reconnection points and current sheets.

This article is organized as follows. In \refsec{sec:grffe}, we briefly review
Maxwell's equations in general relativity and introduce the formulation we adopt
in this work. We also discuss our strategy for maintaining the force-free
conditions in simulations. In \refsec{sec:implementation}, we describe the
numerical implementation of spatial discretization, time stepping, and the
discontinuous Galerkin-finite difference hybrid solver. We present results from
a set of test problems in \refsec{sec:results}, and conclude with a discussion
of result in \refsec{sec:conclusion}.

In this paper, we adopt geometrized ($c=G=1$) Heaviside-Lorentz units, where
electric and magnetic fields have been rescaled by $1/\sqrt{4\pi}$ compared to
Gaussian units. We use the abstract index notation using latin indices ($a, b,
\cdots$) for spacetime tensors, but reserve $\{i,j,k,\ldots\}$ for spatial
tensors. We follow the sign convention of the Levi-Civita tensor from
\cite{Misner1973}, $ \varepsilon_{abcd}  = \sqrt{-g} \, [abcd] $, where $g$ is
the determinant of spacetime metric and $[abcd]=\pm 1$ with $[0123]=+1$ is the
flat-space antisymmetric symbol\footnote{Note that an opposite sign convention
is sometimes adopted in the literature \cite[e.g.][]{Palenzuela2010a,
Palenzuela2013}}.

\section{General relativistic force-free electrodynamics}
\label{sec:grffe}

We begin by outlining the mathematical description used to numerically study
magnetospheric dynamics. This includes a general relativistic formulation of
electrodynamics in a curved spacetime, which is then specialized to the
force-free case: general relativistic force-free electrodynamics (GRFFE).

\subsection{Electrodynamics}
\label{sec:maxwell equations}

The dynamics of electric and magnetic fields is governed by the Maxwell
equations. In covariant form, they are given by
\begin{align}
    \nabla_b F^{ab} & = \mathcal{J}^a \label{eq:Fab} \\
    \nabla_b \dual F^{ab} & = 0 \label{eq:Fdual}
\end{align}
where $F^{ab}$ and $\dual F^{ab} = \varepsilon^{abcd} F_{cd}/2$ are the
electromagnetic field tensor and its dual, and $\mathcal{J}^a$ is the electric
4-current density.

For the standard 3+1 decomposition of the spacetime metric
\begin{equation}
    ds^2 = - \alpha^2 dt^2
        + \gamma_{ij} (dx^i + \beta^i dt) (dx^j + \beta^j dt) ,
\end{equation}
where $\alpha$ is lapse, $\beta^i$ is the shift vector, and $\gamma_{ij}$ is the
spatial metric, the normal to spatial hypersurfaces is given by
\begin{equation}
    n^a = (1/\alpha, -\beta^i/\alpha), 
    \quad
    n_a = (-\alpha, 0) .
\end{equation}

In terms of the normal vector $n^a$, electromagnetic field tensor $F^{ab}$ and
its dual $\dual F^{ab}$ can be decomposed as
\begin{align}
    F^{ab} &= n^a E^b - n^b E^a  - \varepsilon^{abcd}B_c n_d , \\
    \dual F^{ab} &= -n^a B^b + n^b B^a - \varepsilon^{abcd} E_c n_d ,
\end{align}
where
\begin{equation}
    n_a E^a = n_a B^a = 0.
\end{equation}
$E^a = (0, E^i)$ and $B^a = (0, B^i)$ are electric and magnetic fields in the
frame of an Eulerian observer. One can read off $E^a$ and $B^a$ from $F^{ab}$
using the following relations
\begin{align}
    E^a & = F^{ab} n_b , \\
    B^a & = -\frac{1}{2}\varepsilon^{abcd} n_b F_{cd}
    = - \dual F^{ab}n_b\,.
\end{align}

While analytically complete, Maxwell equations cannot be directly evolved
numerically, as any violation of the divergence constraints (Eqs.~\eqref{eq:Fab}
and \eqref{eq:Fdual} with $a=0$) will break strong hyperbolicity of the system
\cite{Schoepe:2017cvt,Hilditch:2018jvf}. This can be avoided by either using
constrained transport approaches \cite{1988ApJ...332..659E} or extending the
system using effective Lagrange multipliers \cite{2002JCoPh.175..645D}. We here
adopt the latter approach. The extended (or augmented) Maxwell equations
\citep{Komissarov2007, Palenzuela2013} are
\begin{equation}\label{eq:extended system - inhomogeneous}
    \nabla_a (F^{ab} + g^{ab}\psi) = -\mathcal{J}^b + \kappa_\psi n^b \psi
\end{equation}
\begin{equation}\label{eq:extended system - homogeneous}
    \nabla_a (\tensor[^*]{F}{^a^b} + g^{ab}\phi) = \kappa_\phi n^b \phi
\end{equation}
\begin{equation}\label{eq:extended system - charge conservation}
    \nabla_a \mathcal{J}^a = 0
\end{equation}
where auxiliary scalar fields $\psi$ and $\phi$ propagate divergence constraint
violations of electric field and magnetic field. $\kappa_\psi$ and $\kappa_\phi$
are damping constants, leading to an exponential damping of the constraints in
the characteristic timescales $\kappa_{\psi,\phi}^{-1}$.

Performing a standard 3+1 decomposition of the extended Maxwell's equations
\eqref{eq:extended system - inhomogeneous}-\eqref{eq:extended system - charge
conservation} using the normal vector $n^a$ and the spatial projection operator
$\tensor{h}{^a_b}\equiv\tensor{\delta}{^a_b} + n^a n_b$, we get
\begin{subequations}
\label{eq:extended maxwell equations}
\begin{equation}
    (\partial_t - \mathcal{L}_\beta)E^i
    - \spatiallc{ijk} D_j(\alpha B_k) + \alpha \gamma^{ij}D_j \psi 
        = -\alpha J^i + \alpha K E^i ,
\end{equation}
\begin{equation}
    (\partial_t - \mathcal{L}_\beta)B^i
        + \spatiallc{ijk} D_j(\alpha E_k) + \alpha \gamma^{ij}D_j \phi
    = \alpha K B^i ,
\end{equation}
\begin{equation}
    (\partial_t - \mathcal{L}_\beta) \psi + \alpha D_iE^i
        = - \alpha \kappa_\psi \psi + \alpha q ,
\end{equation}
\begin{equation}
    (\partial_t - \mathcal{L}_\beta) \phi +\alpha D_iB^i
        = - \alpha \kappa_\phi \phi ,
\end{equation}
\begin{equation}
    (\partial_t - \mathcal{L}_\beta) q + D_i(\alpha J^i) = \alpha q K ,  
\end{equation}
\end{subequations}
where $D_i = \tensor{h}{^a_i}\nabla_a$ is the spatial covariant derivative, $K$
is the trace of extrinsic curvature, $q = -n_\mu \mathcal{J}^\mu$ is the
electric charge density measured by an Eulerian observer, and $J^i =
\tensor{h}{^i_a}\mathcal{J}^a$ is the spatial electric current density. Here we
also defined the spatial Levi-Civita tensor associated with the spatial metric
as
\begin{equation} 
\begin{split}
    \spatiallc{abc} & \equiv n_d\varepsilon^{dabc}\,.
\end{split}
\end{equation}
The Lie derivative along the shift vector applied to a spatial vector $E^i$ is
\begin{equation*}
    \mathcal{L}_\beta E^i = \beta^j \partial_j E^i - E^j \partial_j \beta^i,
\end{equation*} and same for $B^i$ on the left hand side of Eq.~\eqref{eq:extended maxwell equations}, while it is simply a directional derivative (e.g. $\mathcal{L}_\beta (q) = \beta^i \partial_i q$) when applied to a scalar variable.

Evolution equations \eqref{eq:extended maxwell equations} can be cast into
conservative form
\begin{equation}\label{eq:conservative form}
    \partial_t \mathbf{U} + \partial_j \mathbf{F}^j  = \mathbf{S} ,
\end{equation}
with evolved variables
\begin{equation}\label{eq:evolved variables}
  \mathbf{U}
  = \sqrt{\gamma}\left[\,\begin{matrix}
    E^i \\[.5ex]
    B^i \\[.5ex]
    \psi \\[.5ex]
    \phi \\[.5ex]
    q \\
  \end{matrix}\,\right]
  \equiv \left[\,\begin{matrix}
    \tilde{E}^i \\[.5ex]
    \tilde{B}^i \\[.5ex]
    \tilde{\psi} \\[.5ex]
    \tilde{\phi} \\[.5ex]
    \tilde{q} \\
  \end{matrix}\,\right] ,
\end{equation}
fluxes
\begin{equation}\label{eq:fluxes}
\mathbf{F}^j = \left[\,\begin{matrix}
    -\beta^j\tilde{E}^i + \alpha(\gamma^{ij}\tilde{\psi}
        - \spatiallc{ijk} \tilde{B}_k) \\[1ex]
    -\beta^j\tilde{B}^i + \alpha(\gamma^{ij}\tilde{\phi}
        + \spatiallc{ijk} \tilde{E}_k) \\[1ex]
    -\beta^j \tilde{\psi} + \alpha \tilde{E}^j \\[1ex]
    -\beta^j \tilde{\phi} + \alpha \tilde{B}^j \\[1ex]
    \tilde{J}^j - \beta^j \tilde{q} \\
\end{matrix}\,\right] ,
\end{equation}
and source terms
\begin{equation}\label{eq:sources}
\mathbf{S} = \left[\begin{matrix}
    - \alpha\sqrt{\gamma} J^i - \tilde{E}^j\partial_j\beta^i
    + \tilde{\psi}( \gamma^{ij}\partial_j\alpha
        - \alpha\gamma^{jk}\Gamma^i_{jk} ) \\[1ex]
    - \tilde{B}^j\partial_j \beta^i + \tilde{\phi}( \gamma^{ij}\partial_j\alpha
        - \alpha\gamma^{jk}\Gamma^i_{jk} )  \\[1ex]
    \tilde{E}^k \partial_k \alpha + \alpha \tilde{q}
        - \alpha\tilde{\psi}(K + \kappa_\psi) \\[1ex]
    \tilde{B}^k \partial_k \alpha - \alpha\tilde{\phi}(K+\kappa_\phi) \\[1ex]
    0 \\
    \end{matrix}\right],
\end{equation}
where $\Gamma^i_{jk}$ are the Christoffel symbols associated with the spatial
metric. A prescription for the electric current density $J^i$ (Ohm's law) needs
to be supplied to close the system.

\subsection{Force-free limit}
\label{sec:ffe}

In the magnetospheres of neutron stars and black holes, we expect copious
production of electron-positron pairs \cite{Goldreich:1969sb}. The resulting
plasma will be highly conductive, effectively screening electric field
components parallel to the magnetic field. In addition, the magnetization of the
plasma will be very high, allowing us to consider the limit in which the Lorentz
force density vanishes and the plasma becomes force-free.

The force-free conditions are given as
\begin{align}
    F^{ab} \mathcal{J}_b & = 0, \\
    \dual F^{ab} F_{ab} & = 0, \\
    F^{ab} F_{ab} & > 0.
\end{align}

In terms of $E^i$, $B^i$, $q$ and $J^i$, these conditions are
\begin{align}
\label{eq:ff conditions:vanishing lorentz force}
    qE^i + \spatiallc{ijk} J_j B_k &= 0 , \\
\label{eq:ff conditions:edotb is zero}
    E^aB_a = E^iB_i &= 0 , \\
\label{eq:ff conditions:magnetic dominance}
    B^2-E^2 & > 0 ,
\end{align}
where $E^2 = E_aE^a = E_iE^i$ and $B^2 = B_aB^a = B_iB^i$.
The first condition \eqref{eq:ff conditions:vanishing lorentz force} corresponds
to the vanishing Lorentz force density, and the second one \eqref{eq:ff
conditions:edotb is zero} shows the screening of electric field along magnetic
field lines. The third condition \eqref{eq:ff conditions:magnetic dominance} is
called magnetic dominance, and violation of this constraint flags the breakdown
of force-free electrodynamics; characteristic speeds associated with Alfv\'en
modes become complex and Maxwell equations are no longer hyperbolic
\cite{Pfeiffer2013arXiv}. Physically, $E^2 \approx B^2$ means that the plasma
drift speed approaches the speed of light, beyond which the FFE approximation
breaks down.

The force-free conditions also give constraints on the electric current density.
Eq.~\eqref{eq:ff conditions:vanishing lorentz force} gives $J^i$ in the form
\begin{equation} \label{eq:current density with parallel current undetermined}
    J^i = q \frac{\spatiallc{ijk} E_j B_k}{B^2} + \frac{(J_lB^l)}{B^2}B^i ,
\end{equation}
which leaves the parallel component $J_lB^l$ undetermined. The first term on the
right hand side of \eqref{eq:current density with parallel current
undetermined}, the drift current, is perpendicular to both electric and magnetic
fields and shows that electric charge moves collectively with the drift velocity
$v_d = \spatiallc{ijk} E_j B_k / B^2$.

Requiring Eq.~\eqref{eq:ff conditions:edotb is zero} to always be satisfied, we
obtain a closed form expression of the parallel current $J_lB^l$ as
\cite{McKinney2006,Paschalidis2013}
\begin{equation} \label{eq:analytic parallel current}
   J_lB^l = \epsilon_{(3)}^{ijk}\left(
       B_i D_j B_k - E_i D_j E_k\right) - 2 E^iB^j K_{ij},
\end{equation}
which reduces to
\begin{equation}
   J_lB^l = B_i (\nabla \times B)^i - E_j (\nabla \times E)^j
\end{equation}
in the special relativistic limit \cite{Gruzinov1999}.

The parallel current Eq.~\eqref{eq:analytic parallel current} contains the
spatial derivatives of $E$ and $B$, the dynamical variables that we evolve.
Including these derivatives in the source terms changes the principal part of
the Maxwell PDE system, and the resulting system of equations is not strongly
hyperbolic \cite{Pfeiffer2013arXiv}.

A straightforward way to keep the force-free conditions satisfied in numerical
simulations is to algebraically impose Eq.~\eqref{eq:ff conditions:edotb is
zero} and \eqref{eq:ff conditions:magnetic dominance} in the time evolution
\cite{Spitkovsky2006,Palenzuela2010b,Petri2012,Parfrey2012,Cao2015,
Chen:2020,Mahlmann:2021a}. This commonly employed approach exactly ensures the
force-free conditions, but reduces the numerical accuracy to first-order
convergence in time.

As we aim to implement a higher-order numerical scheme for GRFFE, we consider an
alternative strategy. We adopt the driver term approach first implemented in
\cite{Alic2012,Moesta:2011bn} and applied in later studies
\cite[e.g.][]{Most:2020ami,Most2022}. In this method, a stiff relaxation term is
added to the electric current density $J^i$ to continuously damp the violation
of the force-free conditions. We adopt the following electric current density
prescription \cite{Most2022}
\begin{equation} \label{eq:electric current density}
    J^i = q \frac{\spatiallc{ijk} E_j B_k}{B^2}
    + \eta \left[ \frac{E_jB^j}{B^2}B^i
        + \frac{\mathcal{R}(E^2-B^2)}{B^2} E^i \right] ,
\end{equation}
where $\mathcal{R}(x) \equiv \max (x,0)$ is the rectifier function and $\eta$ is
a relaxation parameter. The parallel current consists of the terms in the square
bracket in Eq.~\eqref{eq:electric current density}, each being proportional to
the violation of the force-free conditions \eqref{eq:ff conditions:edotb is
zero} and \eqref{eq:ff conditions:magnetic dominance}. They are coupled to the
evolution of electric field and drive the solution to the force-free limit with
the characteristic damping time scale $\eta^{-1}$. The limiting case
$\eta\rightarrow\infty$ corresponds to the ideal force-free limit.

A caveat to the FFE simulations with a parallel electric current is that the
energy loss from an Ohmic dissipation $J_i E^i$ is removed out from and no
further tracked in simulations; therefore, total electromagnetic energy is not
conserved\footnote{In a MHD model, it is captured as the same amount of increase
in the internal (thermal) energy of the plasma.}. While numerical dissipation
will also contribute to the energy loss, the amount of energy dissipation in
current sheets (corresponding to the rectifier term in Eq.~\eqref{eq:electric
current density}) dominates, albeit likely at a different rate compared to a
full kinetic reconnection model \cite[e.g.][]{Cerutti:2014ysa,Philippov_2015}.

\section{Numerical implementation}
\label{sec:implementation}

In this section, we describe the details of our numerical scheme and its
implementation. We present our method of spatial discretization in
\refsec{sec:domain decomposition}, time integration in \refsec{sec:time
integration}, and the adaptive discontinuous Galerkin-finite difference hybrid
solver in \refsec{sec:dg-fd hybrid}. Our numerical scheme described here is
implemented in the open source numerical relativity code \spectre
~\cite{spectrecode}.

\subsection{Domain decomposition and spatial discretization}
\label{sec:domain decomposition}

The computational domain typically used in astrophysics or numerical relativity
simulations is simple enough to be decomposed into a set of non-overlapping
deformed cubes. We divide the domain into these deformed cubes, which are called
subdomain elements (hereafter simply \emph{elements}). Neighboring elements
share their boundaries at an element interface between them.

Within each element, a spectral expansion can be performed to represent a field
of interest. We also need to define a prescription for handling boundary
corrections from element interfaces. This family of numerical methods is broadly
called spectral element methods \cite{Kopriva2009}. We choose to adopt the nodal
discontinuous Galerkin discretization \cite{hesthaven2007nodal}, so our approach
is formally referred to as a discontinuous Galerkin spectral element method
(DG-SEM), which is often simply called a discontinuous Galerkin (DG) method.

Each element is mapped to a reference cube spanning $\{\xi^1, \xi^2, \xi^3\}\in
[-1, 1]^3$ in the reference coordinate system $\{\xi^i\}$. A coordinate map
$x^i(\xi^j)$ relates the reference coordinates $\xi^j$ to physical coordinates
$x^i$. A set of collocation points $\{\xi^1_i,\xi^2_j,\xi^3_k\}$ are chosen to
represent the solution
\begin{equation}
    u(\xi) = \sum_{i,j,k} u_{i,j,k} \phi_{i,j,k}(\xi)
\end{equation}
where $u_{i,j,k} = u(\xi^1_i, \xi^2_j, \xi^3_k)$ is the value of the solution at
the collocation point $(\xi^1_i, \xi^2_j, \xi^3_k)$, and $\phi_{i,j,k}(\xi)$ is
the nodal basis function
\begin{equation}
    \phi_{i,j,k}(\xi^1_l, \xi^2_m, \xi^3_n) = \left\{ \begin{matrix}
        1, & \text{for } i=l, j=m, k=n \\[1ex]
        0, & \text{otherwise} \\
    \end{matrix} \right\} .
\end{equation}
We use the tensor product basis
\begin{equation}
    \phi_{i,j,k}(\xi) = l_i(\xi^1)\,l_j(\xi^2)\,l_k(\xi^3)
\end{equation}
where $l_d(x)$ is the 1D Lagrange polynomial interpolating collocation points
along the $d$-th axis. We choose to use an isotropic DG mesh with the same
polynomial degree $N$ for each spatial dimension. The resulting nodal expansion
of the solution is
\begin{equation} \label{eq:nodal expansion}
    u(\xi) = \sum_{i=0}^N \sum_{j=0}^N \sum_{k=0}^N u_{i,j,k} \,
        l_i(\xi^1) l_j(\xi^2) l_k(\xi^3) .
\end{equation}
The solution \eqref{eq:nodal expansion} can be also represented in a modal form
\begin{equation} \label{eq:modal expansion}
    u(\xi) = \sum_{p=0}^N \sum_{q=0}^N \sum_{r=0}^N c_{p,q,r} \,
    L_p(\xi^1) L_q(\xi^2) L_r(\xi^3) .
\end{equation}
where $L_p(x)$ is the Legendre polynomial of degree $p$. See also Ref.
\cite{Teukolsky2016} for a detailed derivation of formulating the DG scheme in a
curved spacetime.

In this article, we denote a scheme using the $N$-th degree polynomial basis
(i.e. $N+1$ collocation points) in each spatial dimension as a \dgpn{N} scheme.
For instance, a \dgpn{5} scheme uses $6^3$ collocation points in each element
and a solution is approximated as a fifth degree polynomial in each spatial
direction. When the solution is smooth, a \dgpn{N} scheme exhibits
$\mathcal{O}(L^{N+1})$ spatial convergence where $L$ is the spatial size of an
element.

We mainly use a \dgpn{5} scheme, although we present results for different DG
orders where necessary. We use the Legendre-Gauss-Lobatto collocation points
with the mass lumping approximation \cite{Teukolsky2015}. For a reduced aliasing
error, an exponential filter is applied to rescale the modal coefficients
$c_{p,q,r}$ in Eq.~\eqref{eq:modal expansion}:
\begin{equation} \label{eq:exponential filter}
    c_{p,q,r} \,\to\, c_{p,q,r}
    \prod_{n=\{p,q,r\}} \exp \left[ - a \left(\frac{n}{N}\right)^{2b}
    \right]
\end{equation}
after every DG time (sub)step. We use $a=36$ and $b=50$, which effectively zeros
only the highest mode ($i=N$) and leaves other modes intact\footnote{We note
that this is a common practice adopted in spectral methods for curing aliasing
and has marginal effects on capturing discontinuities, since typically a Gibbs
phenomenon near a discontinuity excites not only the highest mode but multiple
high modes simultaneously.}. Filtering out the highest mode reduces expected
spatial converge of a \dgpn{N} scheme from $\mathcal{O}(L^{N+1})$ to
$\mathcal{O}(L^{N})$.

\subsection{Time integration}
\label{sec:time integration}

Based on the spatial discretization presented in the previous section, evolution
equations can be integrated over time using the method of lines.

The maximum admissible time step size for a \dgpn{N} scheme is
\cite{Cockburn_2001,Dumbser2014}
\begin{equation} \label{eq:DG CFL condition}
    \Delta t \leq \frac{L}{\lambda_\text{max}(2N+1)} \frac{c}{D}
\end{equation}
where $L$ is the minimum (Cartesian) edge length of an element, $\lambda_{\max}$
is the maximum characteristic speed inside the element, $c$ is a stability
constant specific to a time stepper, which is usually of order
unity\footnote{For example, the classic 4th-order Runge-Kutta method has
$c\approx1.39$ \cite{hairer1993}.}, and $D$ is the number of spatial dimensions.

However, usage of a nontrivial coordinate map $x(\xi)$ and a complex geometry of
elements deforms the spatial distribution of grid points, and an actual upper
bound can differ from Eq.~\eqref{eq:DG CFL condition}. As a practical strategy,
we adopt the following expression 
\begin{equation} \label{eq:spectre DG CFL condition}
    \Delta t = f \frac{(\Delta x)_\text{min}}{\lambda_\text{max}}
    \frac{c}{D}
\end{equation}
for the DG time step size, where $(\Delta x)_\text{min}$ is the minimum grid
spacing between DG collocation points in physical coordinates and $f$ is the CFL
factor.

In order to keep the force-free constraint violations as small as possible
during evolution, we aim to use a large value of the damping coefficient $\eta$
for the driver term in Eq.~\eqref{eq:electric current density}, possibly up to
$\eta \Delta t \gtrsim 10$. This implies that the characteristic time scale of
constraint damping $\eta^{-1}$ is smaller than the time step size, which
introduces stiffness in evolution equations and makes explicit time integration
unstable unless an unreasonably small time step is used.

To address the stiffness from rapid constraint damping, we adopt the
implicit-explicit (IMEX) time stepping technique. In particular, we make use of
the IMEX-SSP3(4,3,3) scheme \cite{Pareschi2005}, which is third order in time.
In this IMEX approach, we evolve all quantities explicitly using a standard
3rd-order Runge-Kutta scheme, and treat only the stiff part of the source terms
\eqref{eq:sources} implicitly. Specifically, in the evolution of electric fields
this requires us to solve the following nonlinear algebraic equation at all
substeps,
\begin{equation}
    E^i = (E^i)^* -  \alpha \eta \Delta t' \left[
        \frac{E_j B^j}{B^2} B^i + \frac{\mathcal{R}\left(E^2
    -B^2\right)}{B^2}E^i \right]
\end{equation}
where $(E^i)^*$ are provided values and $\Delta t'$ is an IMEX-scheme-dependent
corrector step size. When $E^2< B^2$, the solution to this equation is
analytical whereas in general cases we employ a three-dimensional Newton-Raphson
solver with a specific initial guess.

In addition to the stiff electric current, we also apply the IMEX time
integration to the hyperbolic divergence cleaning parts to ensure stability,
\begin{align}
    \psi & = \psi^* - \kappa_\psi \Delta t' \psi, \\
    \phi & = \phi^* - \kappa_\phi \Delta t' \phi,
\end{align}
which are linear equations and have exact analytic inversions.

Because of the simplicity of the implicit equations in our evolution system, the
cost overhead from using an IMEX scheme is less than 5\% of the total runtime.
Being able to use much larger time steps more than compensates for this.

\subsection{The discontinuous Galerkin-finite difference hybrid method}
\label{sec:dg-fd hybrid}

This section describes our implementation of the DG-FD hybrid solver for GRFFE
equations. We closely follow the original implementation of \cite{Deppe2021CQG},
which was designed for GRMHD, with several improvements and adaptations.

\subsubsection{Overview of the algorithm}
\label{sec:overview of dg-fd}

Consider an element performing a time step on the DG grid. After each substep of
the time integrator, the candidate solution is monitored by the \emph{troubled
cell indicator} (TCI) to check if the solution is admissible on the DG grid. If
it is admissible, we continue with the updated solution on the DG grid. If the
candidate solution is inadmissible, the troubled cell indicator is flagged, we
undo the DG substep, project the DG solution onto the sub-element FD grid, then
repeat the substep using the FD solver. Evolution on the FD grid proceeds in a
similar way; after every time step the solution gets monitored by the troubled
cell indicator, which determines whether the solution needs to stay on the FD
grid or it is admissible on the DG grid. If the candidate solution looks
admissible on the DG grid, the solution is projected back to the DG grid and the
evolution proceeds using the DG solver.

An optimal number of sub-element finite-difference grid points for a \dgpn{N}
scheme is $2N+1$ \cite{Dumbser2014}. We follow such a prescription, and an
element with $(N+1)^D$ collocation points on the DG grid is switched to
$(2N+1)^D$ FD cells with a uniform grid spacing $\Delta \xi^i = 2 / (2N+1)$ in
the reference coordinates.

At the code initialization phase, all physical quantities are evaluated on the
FD grid to avoid potential spurious oscillations arising from a spectral
representation of the initial data. Next, each element projects evolved
variables onto the DG grid, then either switches to the DG grid or stays on the
FD grid depending on the decision made by the troubled cell indicator.

The projection algorithm of scalar and tensor quantities between DG and FD grids
is described in detail in \cite{Deppe2021CQG}. We use a general sixth-order
accurate interpolation scheme. Since the scheme is general and does not respect
the physical constraints\footnote{or example, the interpolation scheme between
the DG and FD grids does not strictly preserve the force-free conditions or the
divergence (Gauss) constraints.}, repeated applications (i.e., switching back
and forth between DG and FD too frequently) can introduce spurious errors in the
solution. To suppress this behavior, we design the troubled cell indicator to
apply tighter criteria when switching back from FD to DG grid; see
\refsec{sec:troubled cell indicator}.

\subsubsection{Finite difference solver}

Evolution on the finite-difference grid is performed using a conservative
finite-difference scheme \cite{SHU1988,SHU1989}. For an element using a \dgpn{N}
scheme, we divide the reference coordinate interval $[-1, 1]$ into $2N+1$
finite-difference cells and project a solution from the DG grid onto
cell-centered values $\{U_{\hat{\imath}}\}$. A flux-balanced law
\eqref{eq:conservative form} is discretized as
\begin{equation} \label{eq:fd time derivative}
    \frac{d U_{\hat{\imath}}}{dt}
    + \left(\frac{\partial \xi^k}{\partial x^j}\right)
    \frac{\hat{F}^j_{\hat{\imath}+1/2}
        - \hat{F}^j_{\hat{\imath}-1/2}}{\Delta \xi^k}
    = S(U_{\hat{\imath}})
\end{equation}
where we used hat indices to label FD cells and plain indices to label spatial
directions.

Computation of a numerical flux $\hat{F}^j_{\hat{\imath}+1/2}$ is dimensionally
split, and closely follows that of the ECHO scheme \cite{Del_Zanna_2007} (see
also Ref. \cite{Most2019} for an application to neutron star mergers). At the
left and right sides of the FD cell interface $x_{\hat{\imath}+1/2}$, evolved
variables are reconstructed  using their cell-centered values
$\{U_{\hat{\imath}}\}$. In our implementation, densitized electric current
density $\tilde{J}^i$ is also reconstructed to compute fluxes associated with
$\tilde{q}$ (see also Ref. \cite{Palenzuela2013}).

Once face-centered values $U_{\hat{\imath}+1/2}^{L,R}$ are reconstructed, the
interface Riemann flux $F^*_{\hat{\imath}+1/2}$ is computed using the Rusanov
(local-Lax-Friedrichs) flux formula \cite{Rusanov1962}. Since the principal part
of our equations is linear, this solver will reduce to the exact solution (see
Ref. \cite{2002JCoPh.175..645D}).

In order to achieve high-order accuracy, a high-order derivative corrector is
added to the interface Riemann flux to obtain the final numerical flux:
\begin{equation} \label{eq:higher order flux correction}
    \hat{F}_{\hat{\imath}+1/2} = F^*_{\hat{\imath}+1/2} - G_{\hat{\imath}+1/2}^{(4)} .
\end{equation}
The original ECHO scheme uses the Riemann fluxes from cell interfaces (e.g.
$F^*_{\hat{\imath}\pm 3/2}$) for the higher-order correction term
$G_{\hat{\imath}+1/2}^{(4)}$. Since we do not employ a constrained-transport
algorithm requiring a consistent and fixed stencil, we opt for simpler
cell-centered fluxes (e.g., $F_{\hat{\imath}\pm 1}$) for a more compact stencil
and reduced amount of data communications (see Ref.
\cite{Nonomura2013,Chen2016}).

For the simulations presented in this work, we use the WENO5-Z reconstruction
with the nonlinear weight exponent $q=2$ \cite{Borges2008}. The high-order
finite-difference corrector is currently implemented only on Cartesian meshes.
We therefore use it for all of our one-dimensional test problems, where we
assess numerical convergence of the scheme, and defer to future work its
applications in multi-dimensional contexts. Consistent with previous
assessments, we find it sufficient to use only a fourth-order accurate
derivative correction when combined with WENO5-Z \cite{Most2019}.

\subsubsection{Troubled cell indicator}
\label{sec:troubled cell indicator}

In order to decide when to switch between DG and FD grids, our numerical scheme
requires a robust criterion to identify regions of non-smoothness. Such an
approach somewhat shares its idea with popular adaptive-mesh-refinement
criteria. These criteria are inherently problem dependent, and an optimal design
of the troubled cell indicator is at the heart of the DG-FD hybrid method.
Requirements on the indicator include
\begin{enumerate}[label=(\roman*)]
\item A relatively low computational cost
\item Early and robust detection of spurious oscillations developing on the DG
    grid.
\item Being unflagged as soon as the oscillation no longer exists, so that
    evolution can be performed by a more efficient DG solver.
\end{enumerate}

A solution approximated with an $N$-th degree polynomial on the DG grid has
nodal and modal representations where $L_p(x)$ is the Legendre polynomial of
degree $p$. Motivated by the idea of the modal shock indicator devised by
\cite{Persson}, we adopt the oscillation detection criterion
\begin{equation} \label{eq:persson tci}
    \sqrt{\frac{\sum_i \hat{u}_i^2}{\sum_i u_i^2}} \,>\,
        (N+1 -M)^{-\alpha},
\end{equation}
where $\hat{u}$ is the solution with the lowest $M$ modes filtered out i.e.
\begin{equation}
    \hat{u}(\xi) = \sum_{p=M}^N \sum_{q=M}^N \sum_{r=M}^N c_{p,q,r} \,
    L_p(\xi^1) L_q(\xi^2) L_r(\xi^3),
\end{equation}
and the summations $\sum_i$ in Eq.~\eqref{eq:persson tci} with the nodal values
$u_i, \hat{u}_i$ are performed over all DG grid points. The exponent $\alpha$ in
the criterion \eqref{eq:persson tci} controls the sensitivity of the indicator.
Since we filter out the highest mode on the DG grid, the troubled cell indicator
needs to use $M\geq 2$. We use $M=3$ for the troubled cell indicator,
effectively monitoring power from the second and third highest modes.
Empirically we find that $M\lesssim\lfloor(N+1)/2\rfloor$ provides robust
detections of discontinuities without the indicator being excessively triggered.
We use $\alpha=4.0$ following Refs. \cite{Deppe2021CQG,Deppe2022}.

To avoid an element switching back and forth between DG and FD grid in an
unnecessarily frequent manner, we use $\alpha' = \alpha + 1$ when an element is
evolving on FD grid. The tighter bound $\alpha'$ ensures an extra smoothness of
solution when the grid is switched back to DG, preventing it from switching
again to FD within only a few time steps.

Depending on the specific type of an evolved system, one may consider additional
physical admissibility criteria (e.g. positivity of the mass density in the case
of hydrodynamics) for the troubled cell indicator. Since the only physical
constraints in our evolution system, the force-free conditions, are handled by
the stiff parallel electric current, we do not impose any physics-motivated
criteria.

In our implementation of the DG-FD hybrid scheme for GRFFE, we adopt only one
criterion for the troubled cell indicator: application of the modal sensor
\eqref{eq:persson tci} to the magnitude of $\tilde{B}^i$. While it looks
somewhat oversimplified that the information of a single scalar quantity is used
for monitoring a system with nine evolution variables $\{\tilde{E}^i,
\tilde{B}^i, \tilde{\psi}, \tilde{\phi}, \tilde{q}\}$, we show in
\refsec{sec:results} that it is capable of detecting troubled elements in a
satisfactory manner.

\subsection{Outer boundary condition}
\label{sec:no incoming poynting boundary condition}

In 3D simulations, the outer boundary of the computational domain is usually
placed far out to avoid spurious boundary effects leaking into the internal
evolution. Still, in order to suppress potential unphysical noise or reflections
at the outer boundary, we implement a no-incoming Poynting flux boundary
condition as follows. The evolved variables at the outer boundary
$\mathbf{U}_\text{out} = (\tilde{E}^i, \tilde{B}^i, \tilde{\psi}, \tilde{\phi},
\tilde{q})_\text{out}$ are prescribed as follows. First, we copy the values of
$\{\tilde{E}^i, \tilde{B}^i, \tilde{q}\}$ from the outermost grid points. Then,
if the Poynting flux is pointing inward, we set $(\tilde{E}^i)_\text{out}$ to
zero. Divergence cleaning scalar fields $(\tilde{\psi})_\text{out}$ and
$(\tilde{\phi})_\text{out}$ are always set to zero\footnote{Normally, the level
of errors associated with the divergence cleaning part $(\psi, \phi)$ is much
smaller than that of the physical variables $(E^i, B^i, q)$. Spurious
reflections, if any, in the divergence cleaning parts are subdominant to
Poynting fluxes transmitting through the outer boundaries.}. On the DG grid,
$\mathbf{U}_\text{out}$ is fed as an external state when computing the boundary
correction terms. On the FD grid, ghost zones are filled with
$\mathbf{U}_\text{out}$ during the FD reconstruction step.

\section{Results}
\label{sec:results}

In this section, we test and assess our implementation of the DG-FD hybrid
method for evolving GRFFE equations with a suite of robust code validation
problems. We perform 1D tests in \refsec{sec:1d tests}, curved spacetime tests
with black holes in \refsec{sec:black hole tests}, and pulsar magnetosphere
tests in \refsec{sec:pulsar tests}. We also discuss accuracy and efficiency
aspects of the DG-FD hybrid method in \refsec{sec:dg-fd performance}.

\subsection{One-dimensional problems}
\label{sec:1d tests}

\begin{table*}
\caption{Simulation setup for 1D tests in \refsec{sec:1d tests}. Grid resolution
    is increased with $n=0$ (Low), $n=1$ (Med), and $n=2$ (High). For the FFE
    breakdown problem, we use $n=3$ as a reference solution. Each resolution, if
    all elements are switched to FD, is equivalent to $352 \times 2^n$
    finite-difference grid points along the $x$ axis.}
    {\renewcommand{\arraystretch}{1.3}
\centering
\begin{tabular}{@{\extracolsep{0.5cm}} llcccc}
    \hline
    & Domain size & DG Grid points & $\eta$ & CFL factor & Time step size
    $(\times 2^{-n})$ \\
\hline 
    Fast wave & $[-0.5, 1.5]\times[-0.1, 0.1]^2$ & $(192 \times 2^{n}) \times
        6^2$ & $10^6$ & 0.3 & $9.22\e{-4}$ \\
    Alfv\'en wave & $[-1.5, 1.5]\times[-0.1, 0.1]^2$ & & & & $1.38\e{-3}$ \\
    FFE breakdown & $[-0.5, 0.5]\times[-0.1, 0.1]^2$ & & & & $4.61\e{-4}$ \\
\hline
\end{tabular}}
\label{tab:1d tests}
\end{table*}

One-dimensional test problems evolve initial data that only has dependence in
the $x$ direction. We use a computational domain consisting of a single element
along the $y$ and $z$ axes, and impose periodic boundary condition on those
directions. Our lowest grid resolution has 32 elements along the $x$ axis,
resulting in 192 DG grid points. To facilitate comparisons with other results
available in the literature, we note that this resolution is equivalent to 352
grid points if all elements are switched to an FD grid. The number of elements
along the $x$ axis is increased by a factor of two to run medium (64 elements)
and high (128 elements) resolutions. Dirichlet boundary conditions are applied
at both ends of the $x$ axis. We use the CFL factor 0.3 and parallel
conductivity $\eta = 10^6$. Simulation setups are summarized in
Table~\ref{tab:1d tests}. Initial conditions for 1D test problems are summarized
in Appendix \ref{app:1d tests ID}.

\subsubsection{Fast wave}

\begin{figure}
\includegraphics[width=\linewidth]{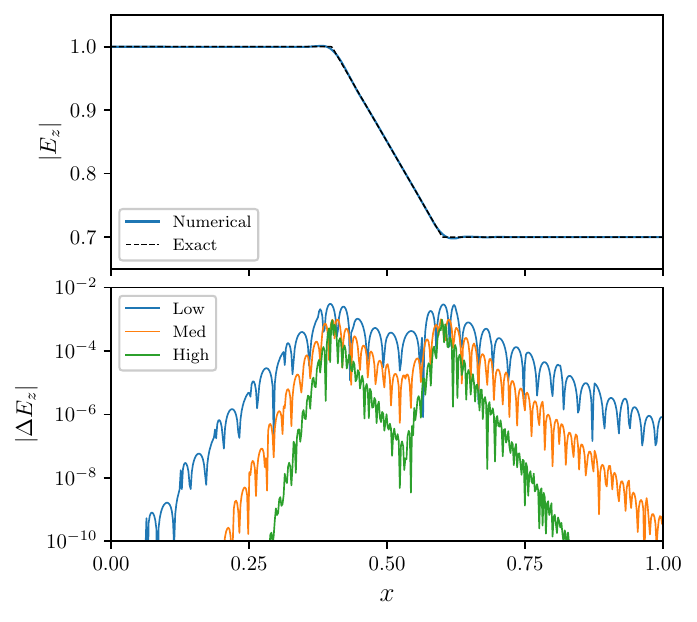}
\caption{Fast wave at $t=0.5$. Top: comparison between the exact solution and a
    numerical solution with the lowest grid resolution. Bottom: error of $E_z$
    for three different grid resolutions.}
\label{fig:test-fastwave}
\end{figure}

Originally due to \cite{Komissarov2002}, this test problem evolves a pure fast
mode propagating in an electrovacuum. The initial profile advects to the $+x$
direction with the wave speed $\mu=1$. The analytic solution is $Q(x,t) = Q(x-t,
0)$ for any physical quantity $Q$.

As shown in Figure \ref{fig:test-fastwave}, our scheme shows good convergence in
flat regions with increasing grid resolution. We observe that the accuracy and
numerical convergence of the solution is substantially lost around two kinks
present in the initial data (corresponding to $x= 0.5 \pm 0.1$ in
\reffig{fig:test-fastwave}) at which spatial derivatives of fields are
discontinuous.

\subsubsection{Alfv\'en wave}
\label{sec:alfven wave test}

\begin{figure}
\includegraphics[width=\linewidth]{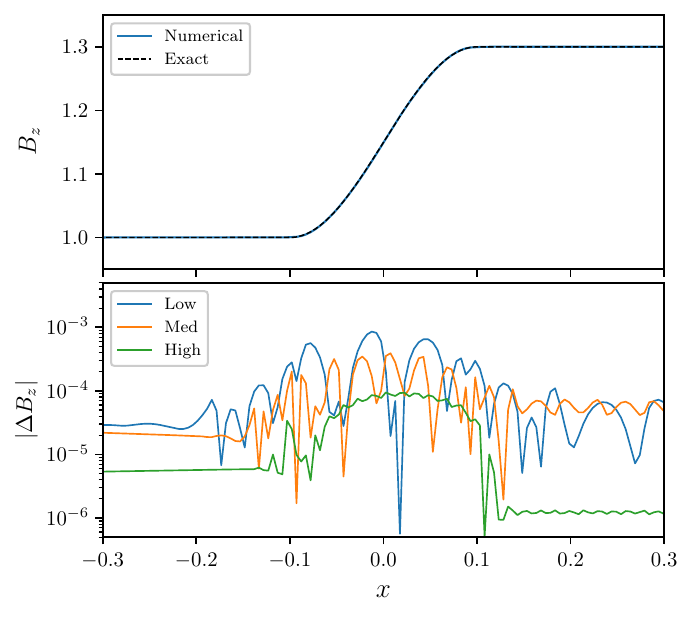}
\caption{Stationary Alfv\'en wave at $t=2.0$. Same plot description as
    \reffig{fig:test-fastwave}.}
\label{fig:test-alfvenwave}
\end{figure}

The stationary Alfv\'en wave problem \cite{Komissarov2004} has a transition
layer $|x| < 0.1$ that maintains a strong parallel current, and the accuracy of
the test results essentially reflects how well a numerical scheme can maintain
the force-free conditions.

We show the result at $t=2.0$ in Figure \ref{fig:test-alfvenwave}. It needs to
be noted that time derivatives of fields at $t=0$, from the initial condition
\eqref{eq:alfven wave initial data}, vanish only if the parallel current $J_i
B^i$ equals Eq.~\eqref{eq:analytic parallel current}. In our approach, the
region $|x| < 0.1$ initially develops a small transient until the stiff
relaxation term becomes fully active within several time steps and effectively
recovers the same value of $J_lB^l$. The amplitude of the initial transient
rapidly decreases at higher grid resolutions. Owing to the higher-order accuracy
of the discontinuous Galerkin discretization, our result shows good convergence
and low amounts of grid dissipation.

\subsubsection{FFE breakdown}

\begin{figure}
\includegraphics[width=\linewidth]{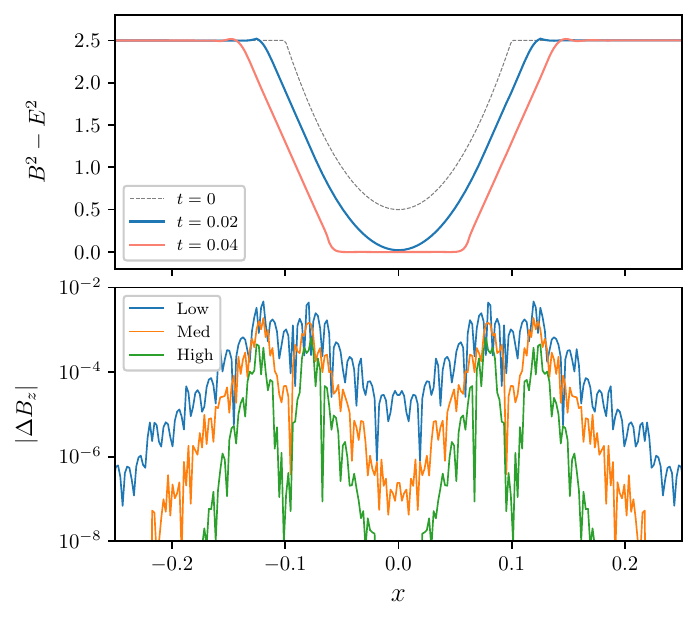}
\caption{FFE breakdown problem. Top: initial data ($t=0$) and a numerical
    solution with the lowest grid resolution at $t=0.02$ and $t=0.04$. Bottom:
    Error of $B_z$ with respect to the reference solution.}
\label{fig:test-ffebreakdown}
\end{figure}

The force-free electrodynamics breakdown problem, originally designed by
\cite{Komissarov2002}, demonstrates that a state initially satisfying the
force-free conditions can later develop into a state violating them. $B^2 - E^2$
decreases over time toward zero in the transition layer $|x|<0.1$ and the
magnetic dominance condition eventually breaks down.

Figure~\ref{fig:test-ffebreakdown} shows numerical results. At $t\gtrsim 0.02$,
the rectifier term restoring the $B^2-E^2 > 0$ condition is switched on and
robustly maintains the magnetic dominance at later times. Since this problem
does not have a closed form solution, we perform an additional higher resolution
run using 256 elements along the $x$ axis and use it as a reference solution to
check the convergence. Similar to the fast wave test, we note the loss of
accuracy and numerical convergence near the kinks present in the solution.

\subsection{Three-dimensional tests: Black hole magnetospheres}	
\label{sec:black hole tests}

\begin{figure}
\includegraphics[width=\linewidth]{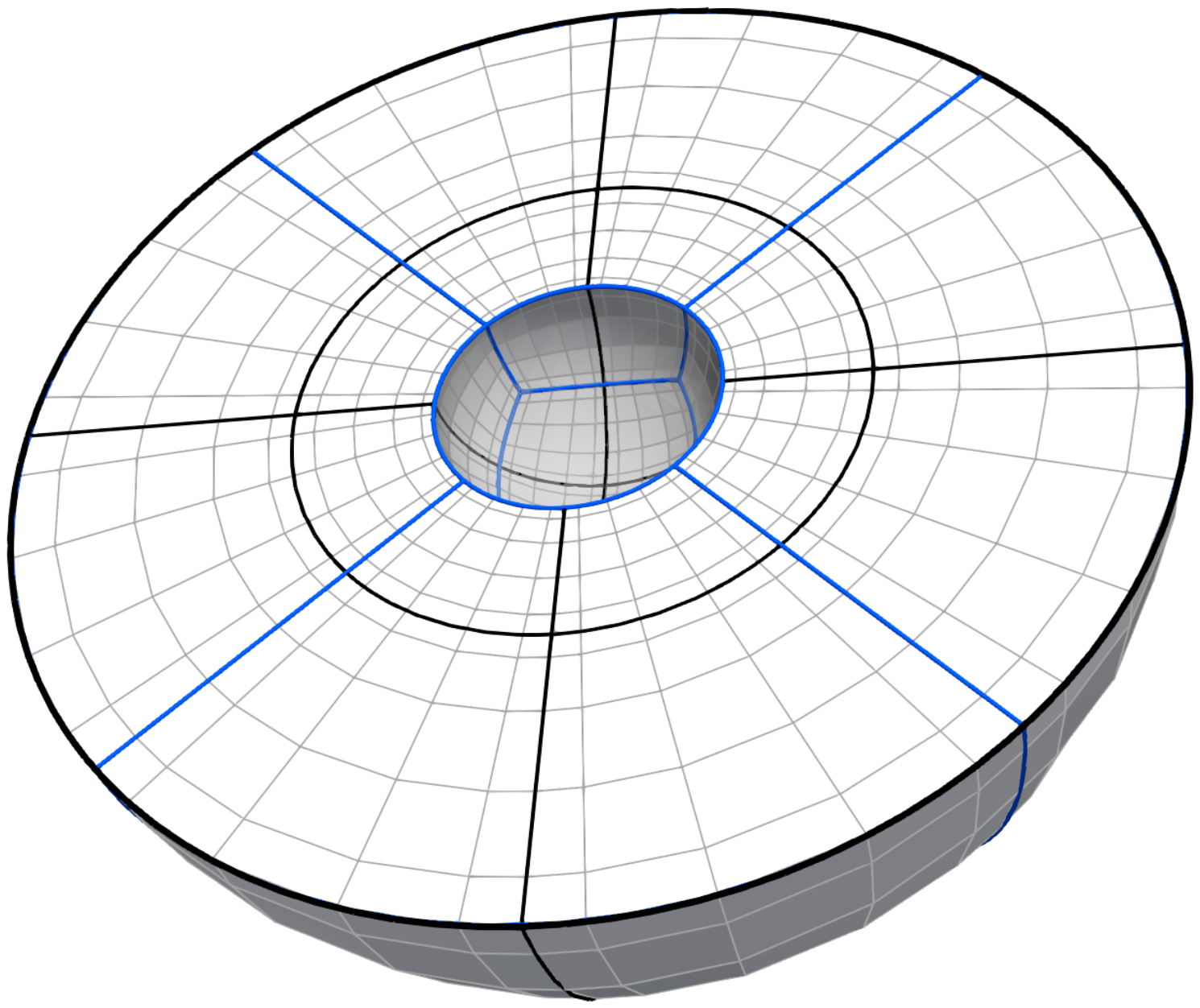}
\caption{A half-cut illustration of the spherical grid used for black hole tests
    in \refsec{sec:black hole tests}. Blue lines show boundaries between the
    cubed-sphere wedges, where black lines show boundaries between each element
    (in this example, there are 8 elements in each wedge). The \dgpn{5} mesh
    consisting of $6^3$ Legendre-Gauss-Lobatto collocation points is shown with
    gray lines. The total number of elements in this example is $N_r \times
    N_\Omega = 2 \times 24$.}
\label{fig:black hole tests grid structure}
\end{figure}

We perform a set of 3D tests in a curved spacetime using black hole
magnetosphere problems. The grid structure of the computational domain is
portrayed in Figure~\ref{fig:black hole tests grid structure}. A spherical shell
spanning the radius $[r_\text{in}, r_\text{out}]$ is split into six cubed-sphere
wedges, which are then further refined into elements. We use an equiangular
coordinate map along angular directions and a logarithmic map along the radial
direction.

\subsubsection{Exact Wald solution}
\label{sec:exact wald solution}

\begin{table*}
\centering
\caption{Convergence tests of different \dgpn{N} schemes on the Wald solution
    (\refsec{sec:exact wald solution}). For each level of grid resolution, we
    show the number of elements used in radial and angular directions, time step
    size $\Delta t / M$, L2 error norm of $\tilde{B}^i$ at $t=5M$, and the
    measured order of numerical convergence.} {
\renewcommand{\arraystretch}{1.2}
\begin{tabular}{@{\extracolsep{0.3cm}} lllcccc}
    \hline
        & Resolution & Elements ($N_r \times N_{\Omega}$)
        & $\Delta t/M$ & Error($\tilde{B}^i$) & Convergence order \\
    \hline
DG-$P_5$ \hspace{1ex} & Low & $1 \times 6$ & $2.90\e{-2}$ & $8.71\e{-4}$ & \\
        & Medium & $2 \times 24$ & $1.15\e{-2}$ & $2.74\e{-5}$ & 4.99 \\
        & High & $4 \times 96$ & $5.76\e{-3}$ & $4.11\e{-8}$ & 4.72 \\
    \hline
DG-$P_7$ & Low & & $1.62\e{-2}$ & $1.23\e{-5}$ & \\    
        & Medium & & $6.29\e{-3}$ & $1.51\e{-7}$ & 6.35 \\
        & High & & $3.14\e{-3}$ & $1.29\e{-9}$ & 6.87 \\
    \hline 
DG-$P_9$ & Low & & $1.03\e{-2}$ & $2.75\e{-7}$ & \\    
        & Medium & & $3.95\e{-3}$ & $2.03\e{-9}$ & 7.08 \\
        & High & & $1.97\e{-3}$ & $8.04\e{-12}$ & 7.98 \\
    \hline
\end{tabular}}
\label{tab:exact wald}
\end{table*}

Wald \cite{Wald:1974} found a stationary electrovacuum solution of Maxwell's
equations in the Kerr spacetime. The solution for the 4-potential is given as
\begin{equation}\label{eq:exact wald solution}
    A_b = \frac{B_0}{2} [ (\partial_\phi)_b + 2a (\partial_t)_b ],
\end{equation}
where $B_0$ is the field amplitude, $\partial_t$ and $\partial_\phi$ are the
Killing vector fields in time and azimuthal directions, and $a = J/M^2$ is the
dimensionless spin of the Kerr black hole.

The Wald solution with $a=0$ satisfies the force-free conditions outside the
horizon. Electric and magnetic fields in Kerr-Schild coordinates are given by
\begin{equation} \label{eq:exact wald initial condition}
\begin{split} 
    \tilde{B}^x & = \tilde{B}^y = 0, \\ 
    \tilde{B}^z & = B_0, \\ 
    \tilde{E}^x & = -\frac{2M B_0 y}{r^2}, \\ 
    \tilde{E}^y & = \frac{2M B_0 x}{r^2}, \\
    \tilde{E}^z & = 0 .
\end{split}
\end{equation}

We evolve the initial condition \eqref{eq:exact wald initial condition} to
$t=5M$ and measure the L2 error norm
\begin{equation}
    L_2(v^i) \equiv \sqrt{\frac{1}{n}
        \sum_{k=1}^n \left[ (v_k^x)^2 + (v_k^y)^2 + (v_k^z)^2 \right]},
\end{equation}
where $v^i = \tilde{B}^i_\text{numerical} - \tilde{B}^i_\text{exact}$ and $n$ is
the number of grid points. The inner domain boundary is placed at $r_\text{in} =
1.99M$, at which no specific boundary condition is imposed. A Dirichlet boundary
condition is imposed at the outer boundary $r_\text{out} = 20M$. Conductivity of
the magnetosphere is turned off by setting $\eta = 0$.

In Table~\ref{tab:exact wald}, we show convergence studies for different orders
of DG schemes $N=5, 7, 9$. Measured convergence of \dgpn{5} and \dgpn{7} schemes
is consistent with the order of DG discretization. A somewhat slower convergence
of the \dgpn{9} scheme can be attributed to other limiting factors such as the
truncation error from time integration or the sixth-order interpolation from the
initial FD grid to DG grid. In all test cases shown in Table~\ref{tab:exact
wald}, all elements stayed on the DG grid throughout the evolution.

\subsubsection{Vacuum Wald problem}
\label{sec:vacuum wald}

\begin{figure*}
\centering
\begin{tabular}{c}
{\includegraphics[width=.46\linewidth,valign=c]{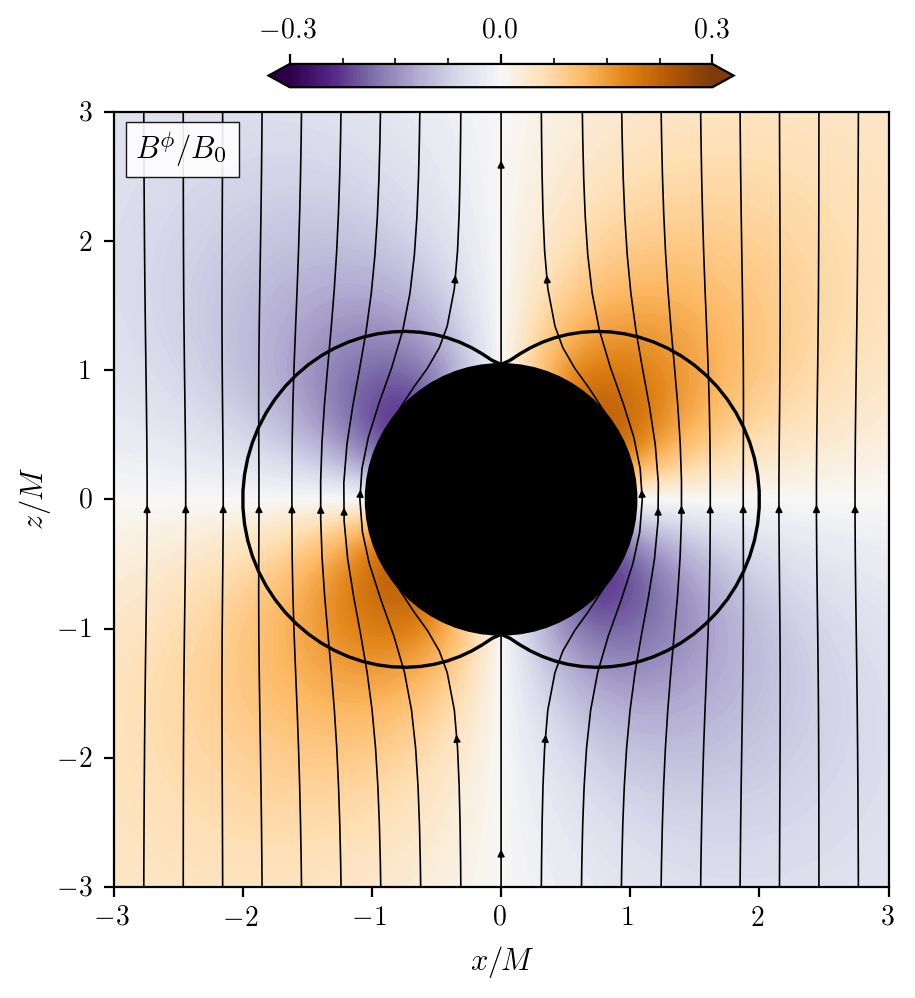}}
\hspace{1em}
{\includegraphics[width=.46\linewidth,valign=c]{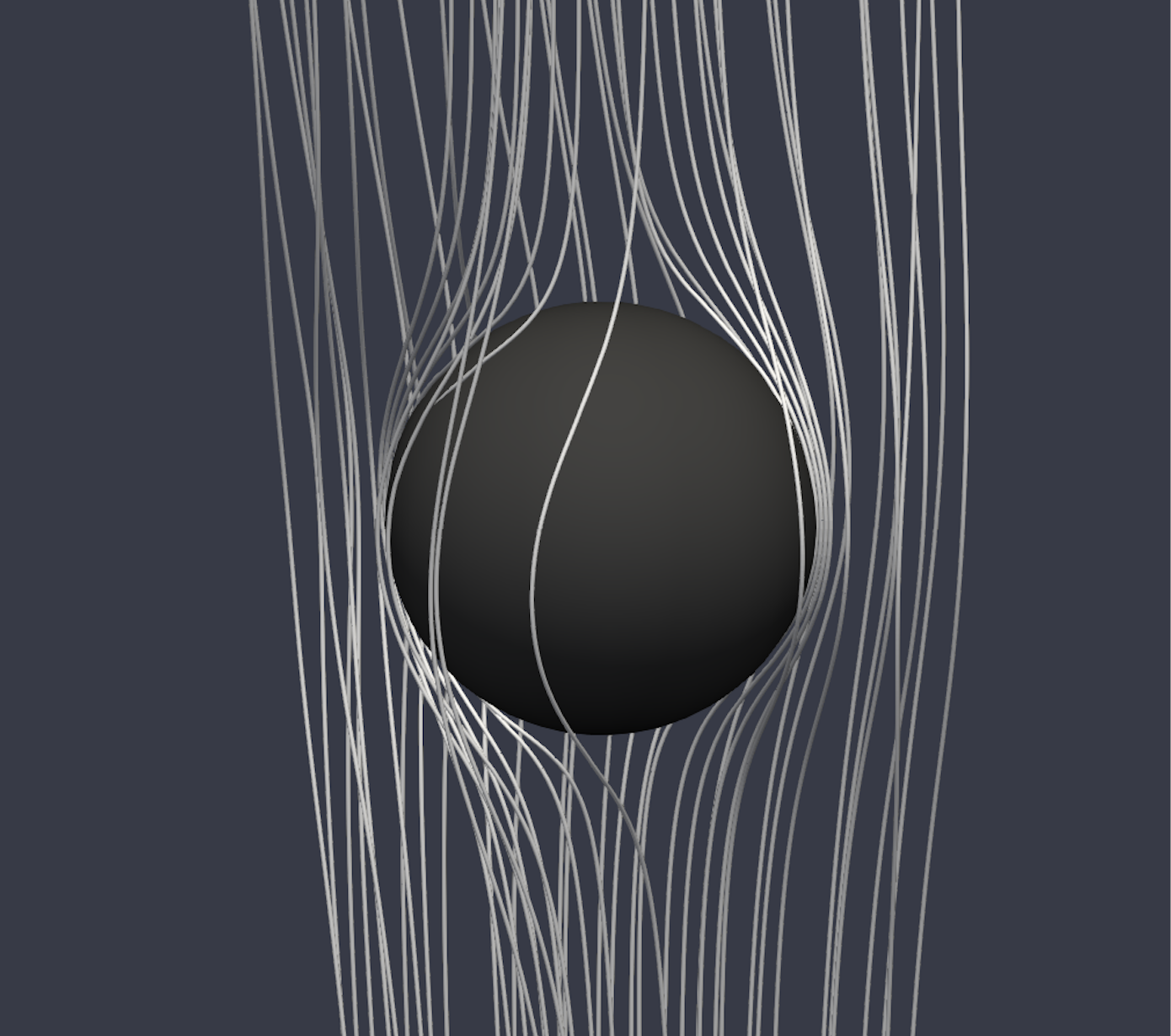}}
\end{tabular}
\caption{Vacuum Wald problem (at $t=125M$) with black hole spin $a=0.999$. Left:
    Toroidal component of the magnetic field and its in-plane field lines on the
    meridional plane. We show the interior of the outer horizon $r=r_+$ with a
    black disk and the ergosphere with black solid lines. Right: A
    three-dimensional visualization illustrates the magnetic field lines (silver
    lines) expelled from the horizon (black sphere).}
\label{fig:vacuum wald}
\end{figure*}

\begin{figure}
\includegraphics[width=\linewidth]{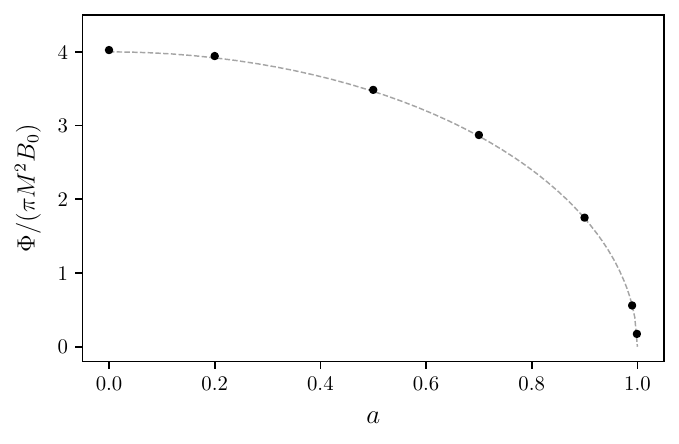}
\caption{Vacuum Wald problem: total magnetic flux through the upper hemisphere
    of the outer horizon versus the spin of the Kerr black hole. Numerical
    results at $t=125M$ (black dots) are shown on top of the analytic prediction
    (dotted line, Eq.~\eqref{eq:vacuum wald magnetic flux}).}
\label{fig:vacuum wald-magnetic flux}
\end{figure}

A time-dependent evolution of electromagnetic fields around a Kerr black hole
can be simulated with the initial magnetic fields given by the Wald solution
\eqref{eq:exact wald solution} where electric fields are set to zero at $t=0$.
The system reaches a steady state that depends on the spin of the black hole and
electrical conductivity of the magnetosphere.

We first simulate the electrovacuum case. The background spacetime is the Kerr
metric with $a=0.999$ in spherical Kerr-Schild coordinates (see Appendices B and
C). The electrical conductivity of the magnetosphere is switched off by setting
$\eta = 0$. We use $N_r \times N_\Omega = 16 \times 96$ elements and use CFL
factor 0.25, resulting in the time step size $\Delta t = 1.97\e{-3}M$. The inner
domain boundary is located at $r_\text{in}=M$, and the no-incoming Poynting flux
boundary condition (see \refsec{sec:no incoming poynting boundary condition}) is
applied at the outer domain boundary $r_\text{out}=125M$.

The evolution reaches a stationary state after $t \gtrsim 80M$. We show the
structure of magnetic fields at $t=125M$ in Figure~\ref{fig:vacuum wald}. The
Kerr black hole expels magnetic field lines, successfully demonstrating the
``Meissner effect'' of black hole electrodynamics \cite{Komissarov:2007rc}.

In a stationary state, total magnetic flux through the upper hemisphere of the
outer horizon has an analytic expression \cite{King1975}
\begin{equation} \label{eq:vacuum wald magnetic flux}
    \Phi = \oint_{r=r_+,\,z>0} B^i d\Sigma_i = \pi r_+^2 B_0
        \left(1 - \frac{a^4}{r_+^4} \right),
\end{equation}
where $a$ is the dimensionless spin of the black hole and $r_+ =
M(1+\sqrt{1-a^2})$ is the outer horizon radius in spherical Kerr-Schild
coordinates. We perform additional simulations varying the black hole spin $a$
using the same grid setup, all reaching stationary states at $t\gtrsim 80M$. We
plot the obtained magnetic flux at $t=125M$ in Figure~\ref{fig:vacuum
wald-magnetic flux}; our numerical results are in an excellent agreement with
the analytic prediction. The troubled cell indicator is flagged at several
innermost elements only for the highly spinning cases with $a \geq 0.90$.

\subsubsection{Magnetospheric Wald problem}
\label{sec:magnetospheric wald}

\begin{figure*}
\includegraphics[width=0.48\linewidth]{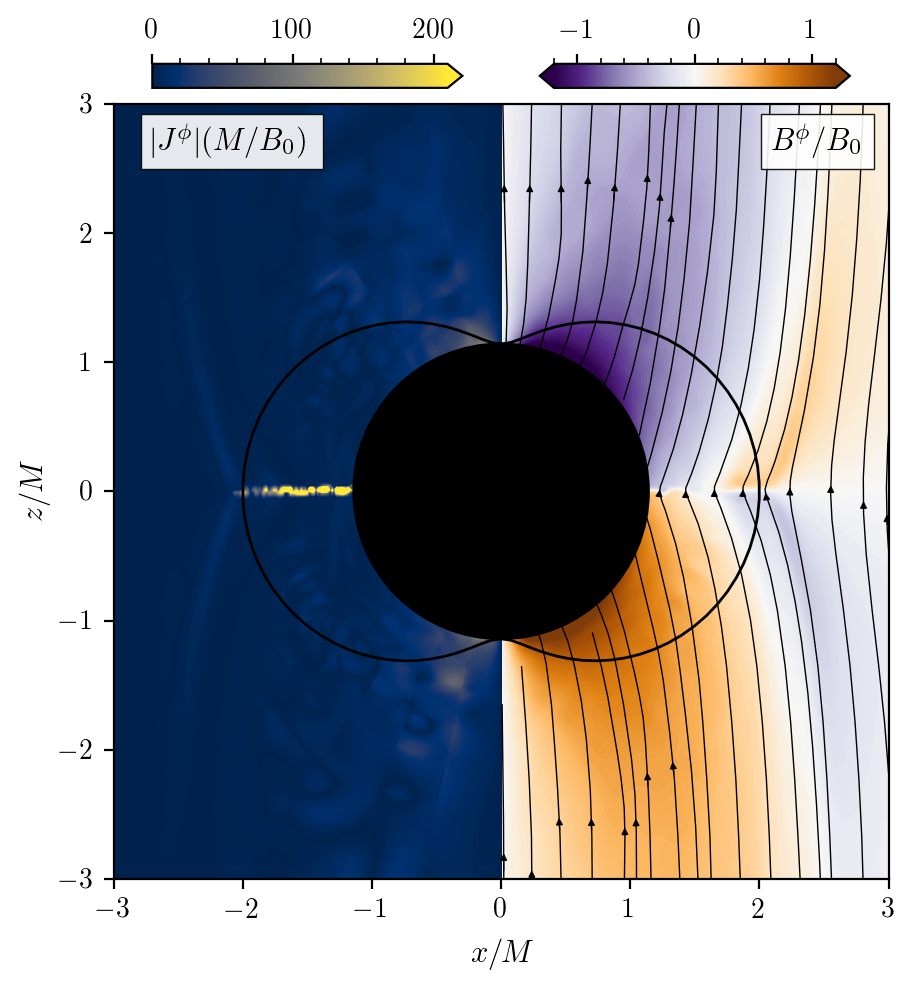}
\hspace{0.3em}
\includegraphics[width=0.48\linewidth]{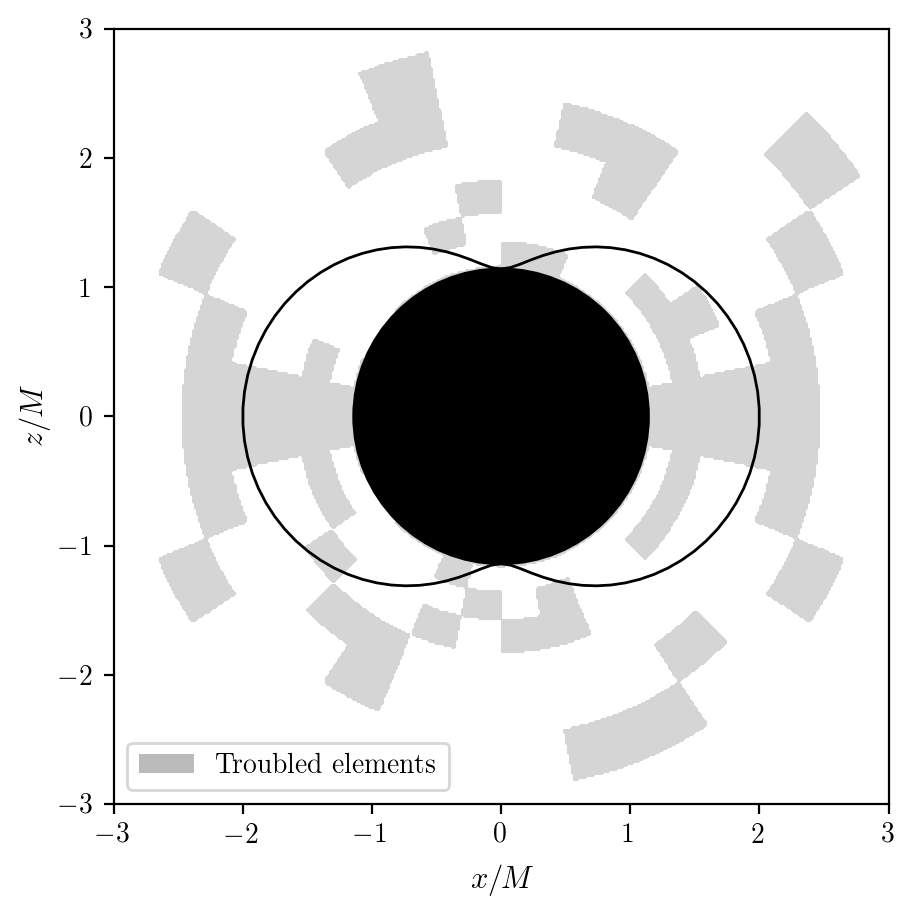}
\caption{Magnetospheric Wald problem at $t=125M$.  In both panels, the interior
    of the outer horizon and the ergosphere are shown with a black disk and a
    thick black line, respectively. Left: toroidal component of the electric
    current density and magnetic field. In-plane magnetic field lines are shown
    with thin black lines in the right half. Right: distribution of troubled
    elements evolved on the FD grid.}
\label{fig:magnetospheric wald}
\end{figure*}

First performed by \cite{Komissarov2004}, this problem models a highly
conductive magnetosphere around a Kerr black hole. The initial condition is the
same as the vacuum Wald problem but now the electrical conductivity of the
magnetosphere is switched on. Compared to the electrovacuum case, the presence
of highly conductive plasma dramatically changes the behavior of the
magnetosphere, since the parallel components of electric fields $E_i B^i$ can be
neutralized by the parallel electric current. There is no analytic solution to
the evolution of this initial value problem, where numerical simulations
\cite[e.g.][]{Komissarov2004,Paschalidis2013,Etienne2017,
Parfrey:2019,Mahlmann:2021a} show that the system reaches a quasisteady state
that resembles the analytically derived solutions of a stationary force-free
magnetosphere \cite{Nathanail:2014aua}.

We perform a test with the black hole spin $a=0.999$ using $N_r \times N_\Omega
= 32 \times 384$ elements with the CFL factor 0.25 ($\Delta t = 9.86\e{-4}M$).
At this grid resolution, if all elements are on the FD grid, there are 176 FD
grid points along the $\theta$ direction with the minimum radial grid spacing
$\Delta r =0.014M$ at the inner boundary $r=M$. Parallel conductivity is set to
$\eta=10^5 M^{-1}$. Small numerical errors and resulting constraint violations
naturally introduce electric charge density into the computational domain via
the parallel current Eq.~\eqref{eq:electric current density}, filling up the
magnetosphere. The system reaches a stationary state at $t \gtrsim 80M$.

We show the result at $t=125M$ in Figure.~\ref{fig:magnetospheric wald}. Inside
the ergosphere, magnetic field lines are dragged by the rotation of the black
hole and a thin current sheet is formed in the equatorial plane. The overall
configuration and topology of the magnetic fields agree well with previous
results reported in the literature. The troubled cell indicator is always
flagged at the elements encompassing the equatorial current sheet, while several
more elements sparsely distributed near the ergosphere are also switched to the
FD grid (right panel of \reffig{fig:magnetospheric wald}).

In the high electrical conductivity limit, magnetic field lines entering the
ergosphere end up crossing the outer horizon \cite{Komissarov2007,Parfrey:2019},
apart from a small portion reconnecting at the equatorial current sheet. Because
of a large grid resistivity in our setup (the ergosphere is radially $\sim$50 FD
grid points across on the equatorial plane), we see that only about half of the
magnetic field lines penetrate the horizon. Some temporal variations of the
current sheet and magnetic field lines near the ergosphere are observed, but the
details of magnetic reconnection and plasmoid formations at the equatorial
current sheet \cite{Parfrey:2019} are not fully resolved at the current grid
resolution.

\subsection{Three-dimensional tests: Pulsar magnetospheres}
\label{sec:pulsar tests}

\begin{figure}
\includegraphics[width=\linewidth]{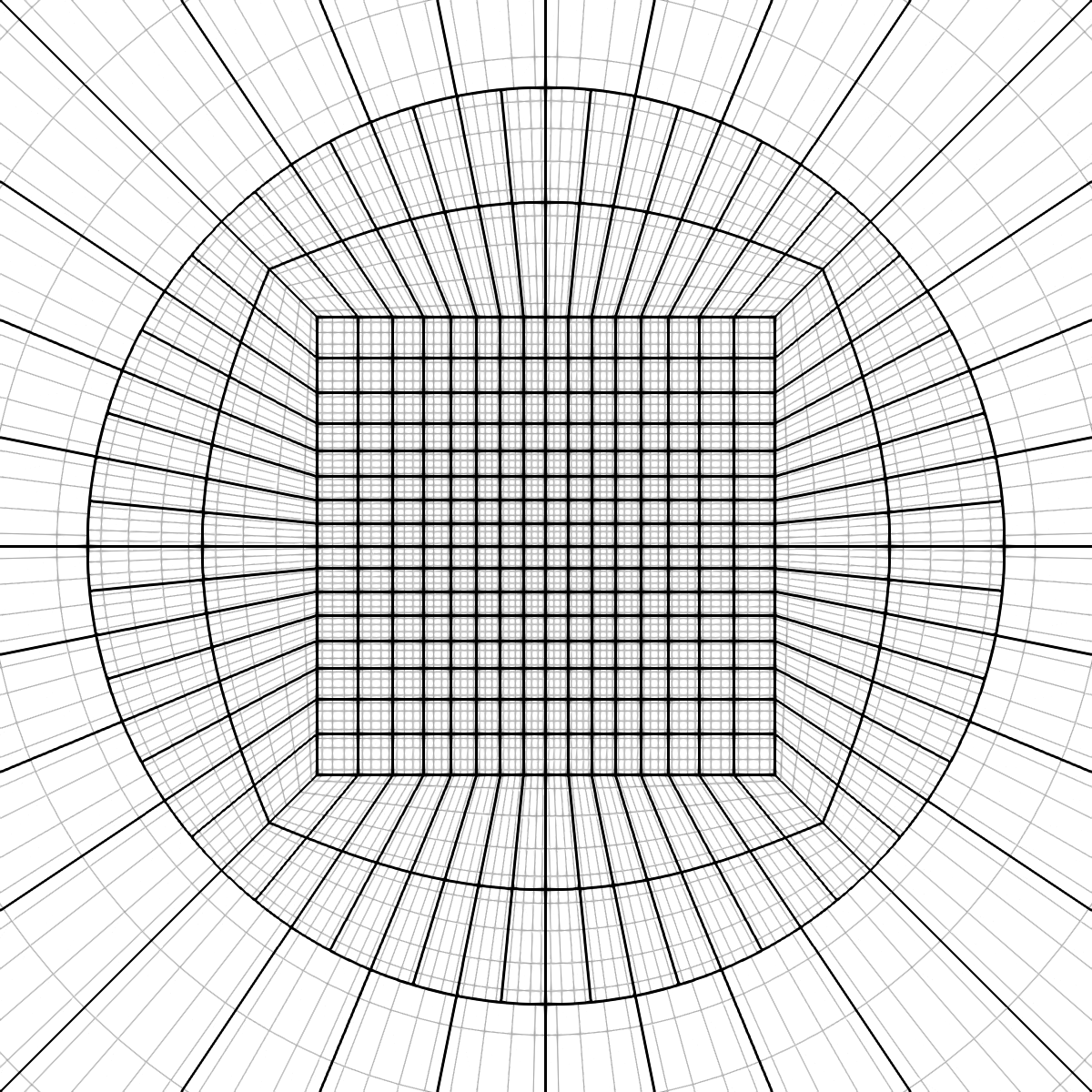}
\caption{A zoom-in view of the computational domain used for the pulsar
    magnetosphere tests in \refsec{sec:pulsar tests}. A sphere domain is divided
    into an inner cube at the center, a layer of cubed-sphere wedges, and an
    outer spherical shell (not fully shown in this figure).}
\label{fig:pulsar domain}
\end{figure}

A conducting sphere threaded with a dipolar magnetic field and rotating in free
space serves as a toy model of pulsars. In the flat spacetime, we set the
initial dipolar magnetic field as
\begin{equation}
    A_\phi = \mu \frac{(x^2+y^2)}{(r^2 + \delta^2)^{3/2}} ,
\end{equation}
where $\mu$ is the magnetic dipole moment, $r^2 = x^2 + y^2 + z^2$, and $\delta$
is a small number to regularize the field at $r=0$. All other variables,
including electric fields, are set to zero everywhere in the initial data.

Rotation of the star is turned on at $t=0$ with a fixed angular velocity $\Omega
\hat{z}$. Inside the star ($r \leq R$), we enforce the perfect conductor
condition 
\begin{equation} \label{eq:ideal mhd condition}
    E^i + \spatiallc{ijk} v_j B_k = 0,
\end{equation}
with the (rigid) rotation velocity field $v^i = \spatiallc{izj}\Omega x_j$. In
practice this is implemented by overwriting electric fields $E^i$ with those
consistent with \eqref{eq:ideal mhd condition} at every substep of time
integration. By this means, the magnetic field is effectively anchored and
corotates within $r\leq R$, whereas fields at $r > R$ are freely evolved. For
consistent behavior of other evolved variables, we also fix $\tilde{\psi}=0$ and
$\tilde{q}=0$ inside the star. The magnetic part of the evolution equations is
freely evolved everywhere. We use $\delta = 0.1R$ for this test.

Denoting the grid refinement level by an integer $l$, the computational domain
consists of an inner cube ($2^{3l}$ elements) and six cubed-sphere wedges
($N_r\times N_\Omega = 2^{l-3} \times 2^{2l}$ elements for each) surrounding it,
wrapped with an outer spherical shell ($N_r\times N_\Omega = 2^{l-3}\times
(6\times 2^{2(l-1)})$). The outer shell uses a logarithmic map along the radial
direction and is fixed to stay on the DG grid. Figure~\ref{fig:pulsar domain}
shows the grid structure for $l=4$. The wedges and the outer shell use an
equiangular map for the angular directions, leading to non-uniform sizes of the
elements in the inner cube: elements closer to the origin have smaller sizes.
Vertices of the inner cube are located at $r_\text{cube} = 10\sqrt{3}R$, and the
cubed-sphere wedges fill the region up to $r_\text{in} = 20R$. The outer shell
extends to the outer domain boundary $r_\text{out} = 60R$, at which the
no-incoming Poynting flux boundary condition is imposed. At the $l=4$ grid
resolution, the total number of grid points is $n_\text{grid}^{1/3}\approx 120$
on the DG grid, and the radius of the rotator $R$ is a single element wide at
the center.

We test with the angular velocity $\Omega = (5R)^{-1}$ and use a parallel
conductivity $\eta = 10^5 R^{-1}$. Our simulation grid is rotated along the $z$
axis with the same angular speed as the rotator.

\subsubsection{Aligned rotator}

\begin{figure*}
\includegraphics[width=\linewidth]{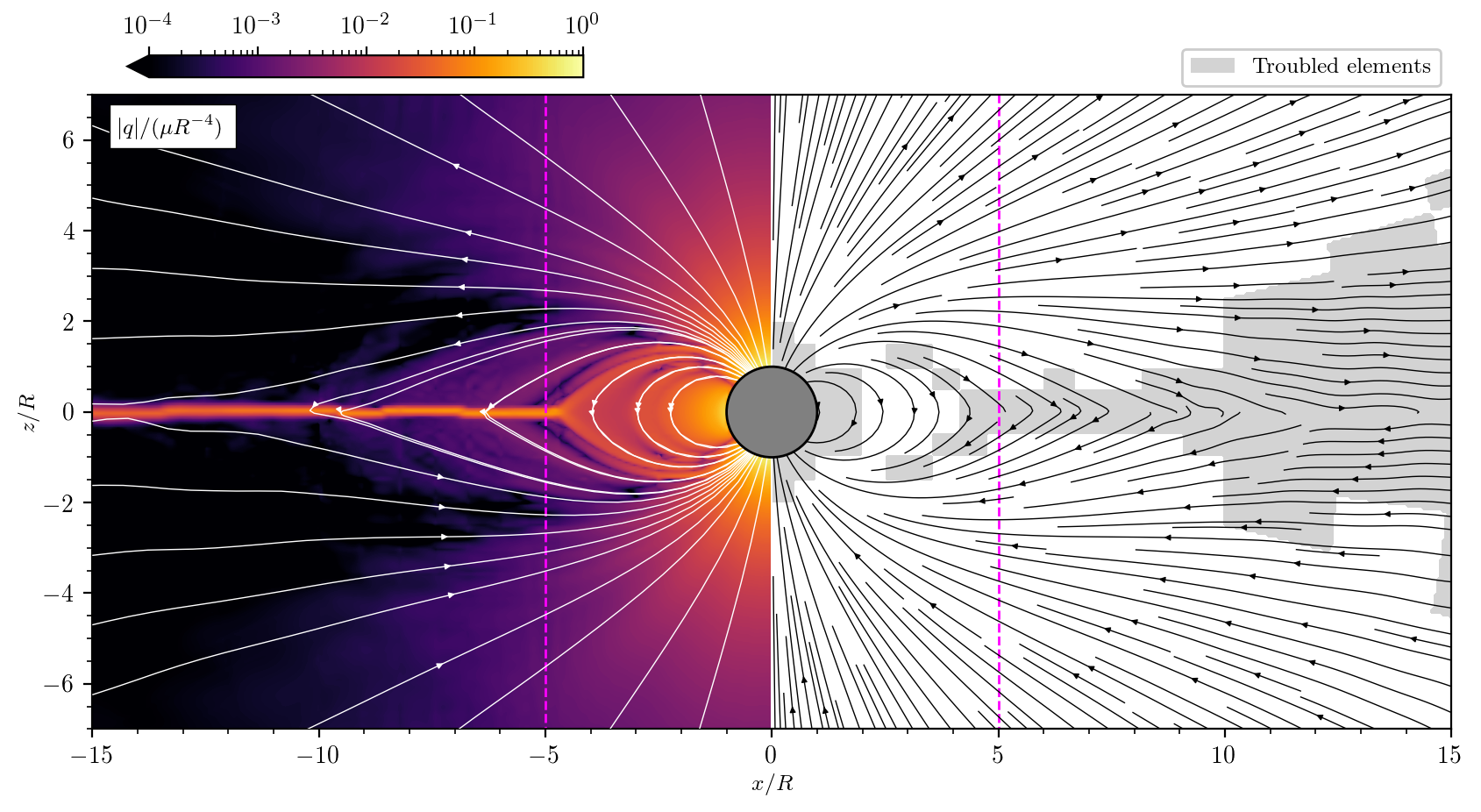}
\caption{Aligned rotator after two rotation periods. Magnetic field lines are
    shown with white solid lines on the left half and black solid lines on the
    right. Electric charge density is shown with a colormap on the left, and the
    distribution of troubled elements is shown with gray shades on the right.
    The light cylinder radius ($r_\text{LC}=\Omega^{-1}$) is shown with magenta
    dashed lines.}
\label{fig:aligned rotator}
\end{figure*}

An aligned rotator ($\theta=0$) is a simple model of a rotating magnetized
neutron star with magnetic moment aligned with the axis of rotation. This
problem is relatively simple because it is axisymmetric, and has been treated in
a large volume of studies (e.g.,
\cite{Goldreich:1969sb,Contopoulos:1999ga,Spitkovsky2006,Komissarov_2006,McKinney:2006sd,Parfrey2012,Cao2015})

We use the $l=5$ resolution, which has 22 FD grid points across the rotator
radius and $n_\text{grid}^{1/3}\approx 240$ total grid points across the DG
grid, along with the CFL factor 0.25 ($\Delta t = 6.03\e{-3} R$). Following an
initial numerical transient, the magnetosphere of the rotator gradually expands
and the system reaches a quasi-steady state after one rotation period.

In Figure~\ref{fig:aligned rotator}, we show the distribution of electric charge
density and the structure of magnetic fields after two rotation periods
($t=20\pi R$). Our scheme successfully reproduces all characteristic features of
the aligned rotator magnetosphere. An equatorial current sheet is formed outside
the light cylinder radius $r_\text{LC} = \Omega^{-1}$, and magnetic field lines
far from the equatorial plane open up to form a monopole-like configuration.
Because of the grid resistivity, magnetic field lines $r \gtrsim 2r_\text{LC}$
spuriously reconnect through the equatorial current sheet. The troubled cell
indicator faithfully tracks the current sheet and the regions with rapid
variations of the magnetic field, which are likely to develop oscillations on
the DG grid, switching elements to the more robust FD grid. The widening of the
distribution of the troubled elements is observed in the outer region $r > 10$.
While elements near the center of the domain has a cubic shape, the elements in
this outer region are deformed (curved) cubes that build up an outer spherical
shell (see \reffig{fig:pulsar domain}), and the Jacobian matrix that maps
logical and physical coordinates is no longer constant within an element.
Ideally, this should have marginal effects on the behavior of the troubled cell
indicator, where empirically we find that the indicator becomes slightly more
sensitive on the elements with curved shapes.

\subsubsection{Oblique rotator}

\begin{figure*}
\begin{tabular}{c}
{\includegraphics[width=.48\linewidth]{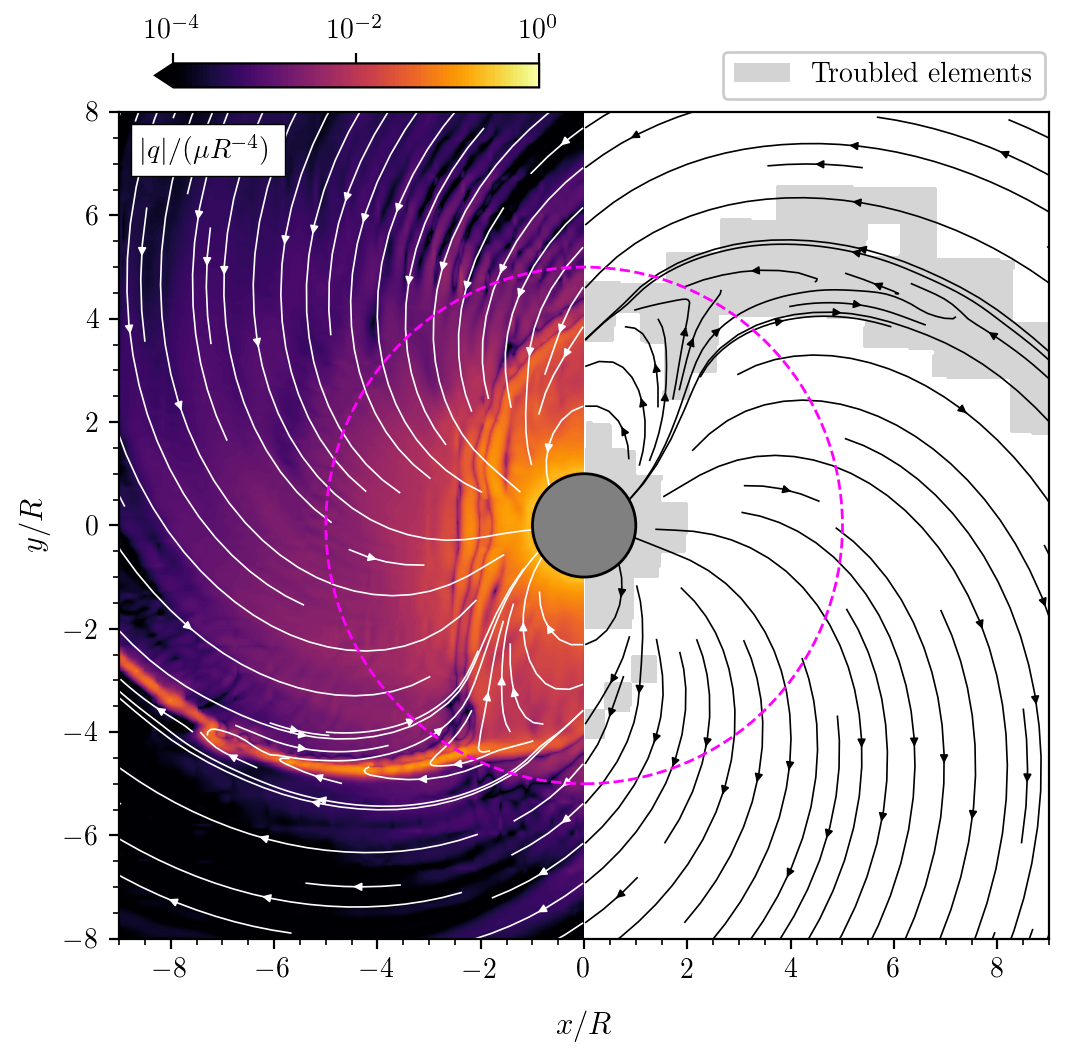}}
\hspace{0.3em}
{\includegraphics[width=.48\linewidth]{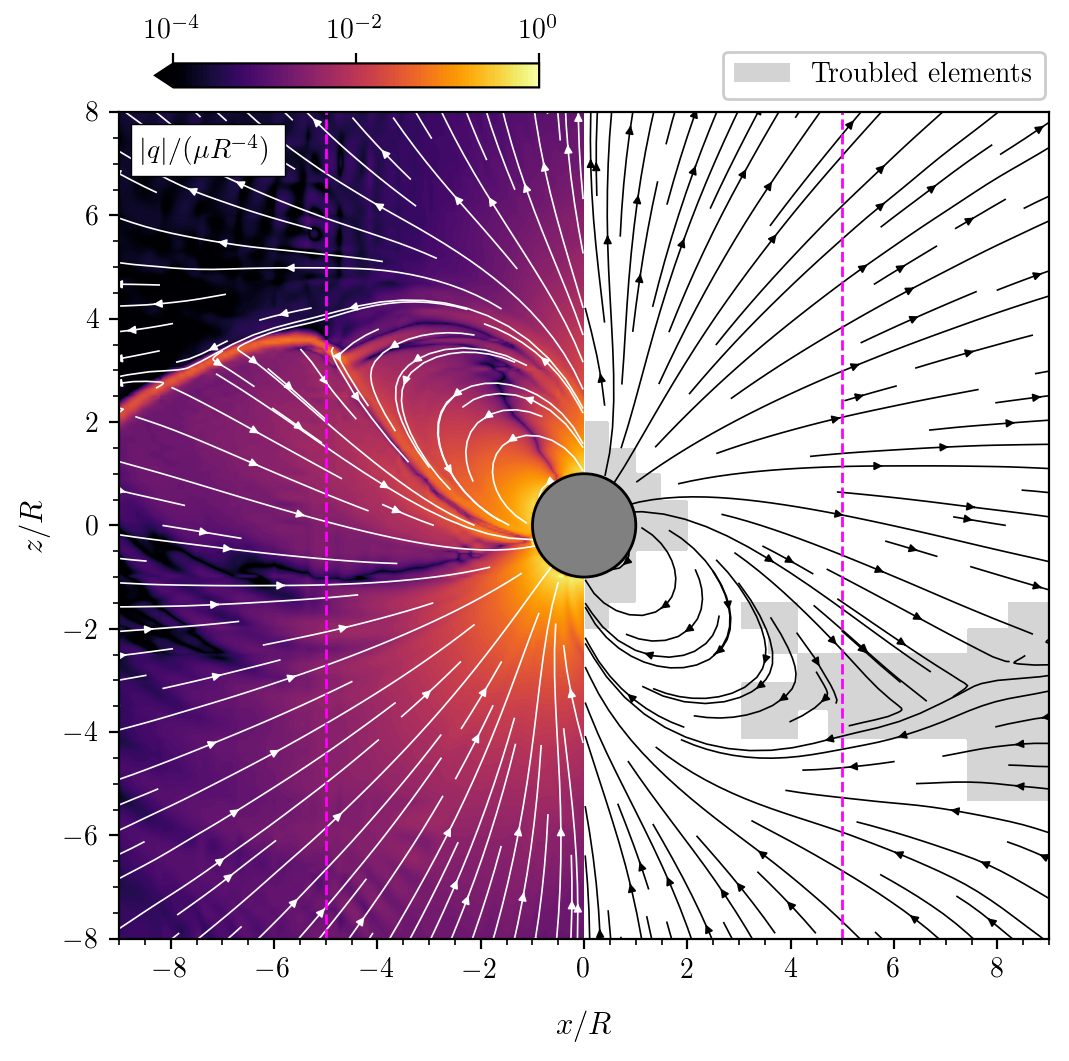}}
\end{tabular}
\caption{Oblique rotator: Simulation snapshot on the equatorial (left) and
    meridional (right) plane after two periods of rotation. Plotted physical
    quantities and their visualizations are the same as \reffig{fig:aligned
    rotator}.}
\label{fig:tilted rotator snapshots}
\end{figure*}

\begin{figure}
\includegraphics[width=\linewidth]{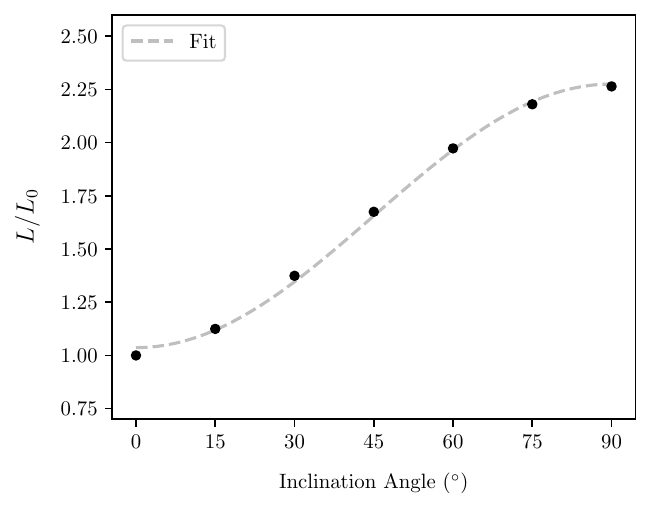}
\caption{Oblique rotator: Inclination angle dependence of the spin-down  
    luminosity. Numerical results (black dots) are fitted with the formula
    \eqref{eq:spin down luminosity angle dependence}, shown with a gray dashed
    line.}
\label{fig:tilted rotator-luminosity-inclination}
\end{figure}

Having a misalignment angle between the magnetic moment and the rotation axis,
an oblique rotator serves as a more realistic model of astrophysical pulsars
\cite{Spitkovsky2006,Kalapotharakos:2008zc,Petri2012,Petri:2016,
Carrasco:2018kdv}. Since the configuration is no longer axisymmetric, a full 3D
simulation is required to study this problem.

We use the same simulation setup as the aligned rotator test, but tilt the
initial magnetic field by an inclination angle $\theta=\pi/4$. The system
reaches a steady state after about one rotation period.

Figure~\ref{fig:tilted rotator snapshots} shows simulation snapshots on the
equatorial plane (left panel) and meridional plane (right panel) after two
periods of rotation. Generic features of the solution are similar to the aligned
rotator. Beyond the light cylinder radius, a current sheet is formed and
magnetic field lines are opened up. Now that the magnetic axis is misaligned
with the rotation axis, the current sheet has a periodically modulated curved 3D
geometry, appearing as a spiral pattern on the equatorial plane. It is clearly
visible that the troubled cell indicator robustly captures and tracks
magnetic reconnection points and the spiral current sheet so that the solution
can be evolved on the FD grid in those regions. The remainder of the domain
keeps evolving on the DG grid, which is computationally more efficient.

One can compute the spin-down luminosity of the rotator
\begin{equation} \label{eq:spin down luminosity}
    L = \oint S_i d\Sigma^i,
\end{equation}
where
\begin{equation}
    S^a = \frac{E^2 + B^2}{2} n^a + \spatiallc{abc}E_b B_c
\end{equation}
is the Poynting vector. We perform simulations with a lower grid resolution
$l=4$ for different inclination angles and compute the spin-down luminosity
Eq.~\eqref{eq:spin down luminosity} after two rotation periods at $r=6R$.
Figure~\ref{fig:tilted rotator-luminosity-inclination} shows the measured
values. The inclination dependence of the spin-down luminosity $L$ is well
fitted with the relation \cite{Spitkovsky2006}
\begin{equation} \label{eq:spin down luminosity angle dependence}
    L = L_0(k_1 + k_2 \sin^2 \theta)
\end{equation}
yielding $k_1 = 1.04$ and $k_2 = 1.24$, where $L_0$ is the luminosity of the
aligned configuration ($\theta=0$).

\subsection{Performance comparison between DG and FD grids}
\label{sec:dg-fd performance}

One of the main goals of this work is to assess the performance and cost-saving
potential of using a DG-FD hybrid method for global FFE simulations of compact
binary magnetospheres. We do so in two steps. First, we establish an accuracy
benchmark to identify corresponding DG and FD resolution requirements for the
same level of accuracy. Second, using this optimal choice, we estimate the
cost-savings/speed-up factor of the DG-FD hybrid methods over traditional FD
approaches for the problems presented in this work.

\subsubsection{Accuracy comparison}

\begin{figure}
\includegraphics[width=\linewidth]{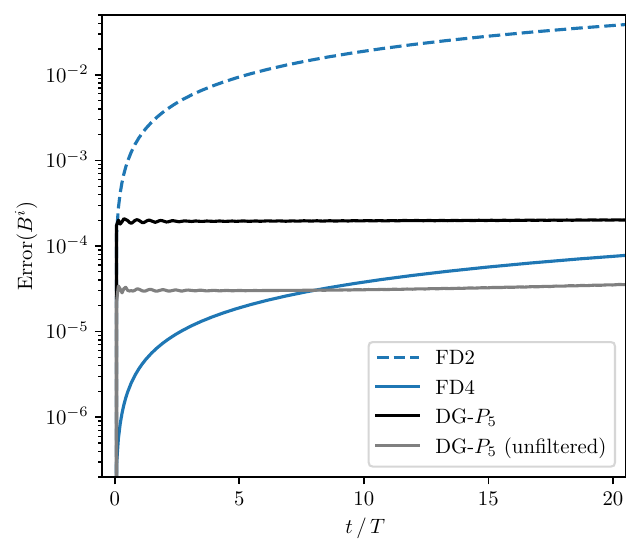}
\caption{Error norm of magnetic field over time from a test evolving a
    sinusoidal fast wave in a periodic box (described in \refsec{sec:dg-fd
    performance}). For the FD runs, we perform the test without (FD2) and with
    (FD4) the high-order flux correction.}
\label{fig:dg-fd accuracy}
\end{figure}

Depending on which scheme is taken as the baseline, the DG-FD hybrid method can
be interpreted either as a sophisticated shock-capturing technique for the DG
method, or an FD method that compresses a group of cells into a high-order
spectral representation on smooth regions \cite{Deppe2022}. The `exchange ratio'
between these two grids, namely $(N+1)^D$ DG grid points and $(2N+1)^D$ FD grid
points, has been determined by equalizing the maximum admissible time step sizes
\cite{Dumbser2014}.

An important follow-up question is comparing the accuracy between the DG and FD
grids when the number of grid points is subject to the above ratio. For example,
does a compression of $11^3$ FD grid points into a $P_5$ DG mesh with $6^3$ grid
points in a smooth region lead to an increase or decrease in accuracy? Clearly,
the answer is highly dependent on the details of DG (e.g. order of the
polynomial, how filtering is applied) and FD solvers (e.g. reconstruction
scheme, high-order corrections), which needs to be assessed on a case-by-case
basis.

However, it is desirable that the DG and FD solvers have similar levels of
accuracy in smooth regions. For instance, coupling a low-order DG scheme with a
very high-order FD reconstruction is not ideal since the evolution on the FD
grid is computationally too expensive considering the overall achievable
accuracy with such a choice. This may possibly make adopting a low-order FD
scheme and using an increased number of elements overall more efficient. On the
other hand, hybridizing a high-order DG scheme with a low-order FD scheme
introduces a relatively large numerical diffusion on the FD grid, artificially
smearing out important features, especially on smooth regions close to a
discontinuity. In this case, the quality of the solution from the DG-FD
hybridization, despite its algorithmic complication, might be no better than
simply applying an aggressive DG limiter.

A desired sweet spot is setting the DG and FD discretization to have the same
order of convergence. As a fiducial case, we consider the same setup used for
the 1D test problems: a \dgpn{5} with the highest mode filtered out and a FD
solver using the WENO5-Z reconstruction with $q=2$ along with the fourth-order
derivative corrector. Both discretizations are fifth-order convergent for smooth
solutions.

We perform a simple numerical experiment as follows. The 1D fast wave problem
(see \refsec{app:1d tests ID}) is modified to a smooth initial profile $E_z =
-B_y = \sin(2\pi x / \lambda)$ with the wavelength $\lambda = 2$. The
computational domain $[0, \lambda]^3$ is split into four elements in each
spatial direction. The initial condition is evolved up to 20 wave crossing times
using periodic boundary condition.

Figure \ref{fig:dg-fd accuracy} shows the time evolution of the error norm of
the magnetic field. While both the DG and FD discretizations used in this test
have the same fifth-order convergence, the FD grid has a twice smaller grid
spacing and shows better accuracy for $t \leq 20 \lambda$. We note the
boundedness of the error norm on the DG grid, demonstrating a lower numerical
dissipation and its resulting suitability for problems involving long-range wave
propagation. By contrast, the error norm on the FD grid increases monotonically
with time, approaching the same level of error as the DG grid at $t\geq
20\lambda$.

We cautiously interpret this result in the following way. In long-term
magnetospheric simulations, replacing the group of $(2N+1)^D$ FD cells with a
\dgpn{N} spectral mesh in smooth regions likely does not harm global accuracy.
In simulations with realistic astrophysical scenarios, the solution is not
smooth everywhere but will have localized large gradients such as current sheets
separated by smooth regions. The global numerical error will then be dominated
by the regions with large gradients, since a shock-capturing FD scheme (such as
WENO5-Z) will fall back to a lower order.

As an additional example, in \reffig{fig:dg-fd accuracy} we also show results
from an unfiltered \dgpn{5} scheme and an FD scheme without the high-order flux
correction. The accuracy of the solution on the FD grid is significantly lower
without the high-order corrections, the error being even higher than the DG grid
using twice fewer grid points per spatial dimension. The unfiltered \dgpn{5} has
a sixth-order convergence and shows smaller error than the FD grid at $t \gtrsim
10\lambda$. As soon as the DG grid has a higher order of discretization than the
order of FD discretization, DG shows a better accuracy in spite of having fewer
grid points.

In summary, in particular for the hybridization of a \dgpn{5} and a WENO5-Z FD
scheme, we conclude that switching from the FD to the DG grid results in a
marginal loss of local accuracy in smooth regions, which is unlikely to affect
the global error in actual simulations.

\subsubsection{Efficiency}

\begin{figure}
\includegraphics[width=\linewidth]{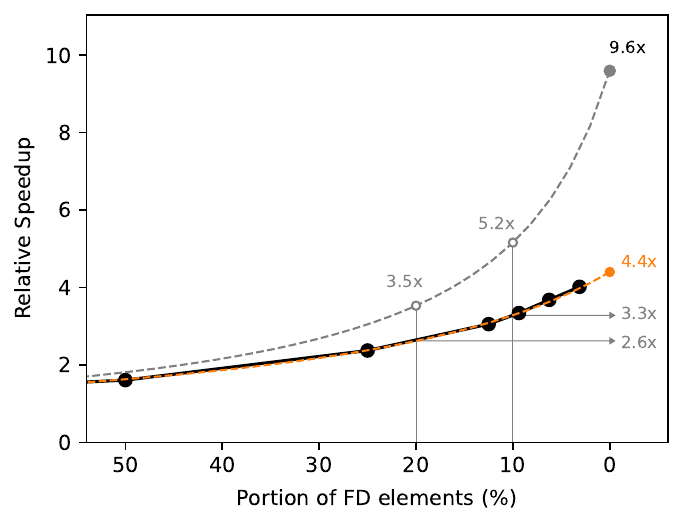}
\caption{Single-core wall-clock speedup of the 1D Alfv\'en wave problem for
    different fractions of elements using the FD grid. An ideal scaling between
    all-DG and all-FD is shown with a dashed gray line. The  measured scaling
    (yellow dashed line) fit shows that a DG element runs 4.4$\times$ faster on
    average than an FD element. 10\%/20\% fraction of FD elements give
    3.3$\times$ (5.2$\times$) / 2.6$\times$ (3.5$\times$) overall (ideal)
    speedup as shown by the solid (dashed) gray line.} 
\label{fig:dg-fd speedup}
\end{figure}

Having confirmed that the DG and FD grids show a comparable level of accuracy in
the setup we use, we now assess potential computational cost savings when using
the hybrid scheme. Since the number of grid points on the DG grid is fewer than
the FD grid by a factor of $(11/6)^3 = 6.2$ in 3D, a similar amount of
computational speedup is naturally anticipated. In order to quantify the actual
speedup in our implementation, we run the stationary Alfv\'en wave test
(\refsec{sec:alfven wave test}) using $(N_x, N_y, N_z) = (8, 32, 8)$ elements.
We manually force elements to stay on the DG grid for $y>0$ and on the FD grid
for $y<0$\footnote{Therefore, in this controlled experiment, the overhead
related to (i) execution of the TCI and (ii) rolling back the time step on
troubled elements are excluded. Each element still sends out ghost-zone data to
neighboring elements at every time step.}. The fraction of elements running on
the FD grid is changed by shifting the upper and lower bounds of the $y$
coordinate. This allows us to vary the fraction of FD to DG grid points in a
controlled way. We additionally disable parallelization and carry out all tests
in this section on a single CPU core to disentangle parallel scaling from
algorithmic performance. Our fiducial benchmark is then given by the overall
wall-clock time of the evolution algorithm.

Figure~\ref{fig:dg-fd speedup} shows the relative speedup compared to the case
when all elements are using the FD grid. Since the DG solver does not involve
computationally expensive reconstruction steps and has less data communication,
it can perform $\sim$50\% more grid point updates per second compared to the FD
solver, which results in a combined 9.6$\times$ overall speedup. In absolute
terms, the measured zone-cycles per CPU second are 108K when all elements are on
the DG grid, and 69K when all elements are on the FD grid.

Assuming perfect scaling, the overall speedup relative to the all-FD case can be
estimated with the  simple formula
\begin{equation}
    \frac{1}{x + (1-x)/f}
\end{equation}
where $x$ is the portion of elements using the FD grid and $f$ is the speedup
factor of the DG grid with respect to the FD grid. In \reffig{fig:dg-fd
speedup}, we show the ideal speedup scaling with $f=9.6$ (gray dashed curve).
However, our measurements show that as soon as there is any portion of FD
elements, the effective speedup factor drops down to $f=4.4$, implying the
presence of an algorithmic bottleneck.
When half the elements are using the FD grid, measured zone-cycles per second
are 65K, even a bit slower than the all-FD case. This somewhat unexpected drop
in performance is likely to be related to an extra interpolation step required
at the interface between DG and FD elements to convert the ghost zone data sent
between the elements. As a representative number, we quote the achievement of
3.3$\times$ (5.2$\times$) / 2.6$\times$ (3.5$\times$) overall (ideal) speedup
when 10\% / 20\% of elements are using the FD grid.

A separate, detailed profiling of the code suggests at most 10\% overhead from
the controlling part of the DG-FD hybrid algorithm (applying the TCI to the
solution and undoing a time step if an element is troubled), which was excluded
in the speedup test described above. Comparing simulations of a smooth wave
solution using the DG grid on all elements with the adaptive DG-FD scheme turned
on and off showed less than 4\% of difference in total runtime.

\section{Conclusions}
\label{sec:conclusion}

We have developed a new numerical scheme for general-relativistic FFE based on a
DG-FD hybrid method. The numerical scheme combines a high-order spatial
discretization with IMEX time stepping to handle stiff source terms associated
with maintaining the FFE constraints. We have further implemented a troubled
cell indicator capable of flagging spurious features in the DG evolution,
allowing the associated elements to transition to a more dissipative
conservative FD scheme. In this way, the scheme achieves high-order convergence
for smooth problems while robustly tracking and capturing large gradients
present in solutions such as current sheets. Our implementation is based on the
open-source \spectre~code and successfully passes and reproduces a suite of
standard test problems in one- and three-dimensions. In particular, we achieve
up to eighth-order numerical convergence in smooth vacuum problems. A
quantitative measure of the numerical resistivity in our scheme, in particular
using the approaches by \cite{Rembiasz2017,Mahlmann:2021b}, will be explored in
future works.

In order to assess potential cost savings of this approach over more traditional
FD-only schemes, we have performed a quantitative assessment of its accuracy and
efficiency. We find that our approach has a potential to speed up FD simulations
by the factor of 2-3 with little to no loss of accuracy. We further demonstrate
an additional optimization potential of (in some cases) up to a factor 2, when
compared to the ideal speed up of the code. Similar or even larger performance
gains have been reported when adopting GPU-based parallelization strategies
\cite[e.g.][]{Liska_2022,DelZanna2024,Grete2021}. Additional improvements may
come from a more optimal set of troubled cell indicator criteria or a dynamic
power monitor \cite[e.g.][]{Szilagyi_2014}, which can potentially facilitate a
more economical grid switching between DG and FD.

The DG-FD hybrid scheme presented here is particularly well suited to study wave
propagation as well as accuracy-limited problems, such as steady-state twists or
magnetospheric explosion dynamics that evolve on long timescales
\cite[e.g.][]{Parfrey:2013gza,Yuan:2019mdb,Yuan:2020eor}. Such studies will be
the subject of future work.

\section{Acknowledgements}

The authors are grateful to Kyle Nelli and Nils Vu for helpful discussions and
technical support during various stages of this project. YK acknowledges
Crist\`obal Armaza for encouraging conversations.

Computations were performed on the Wheeler cluster and Resnick HPC Center at
Caltech.  ERM acknowledges support by the National Science Foundation under
Grant No. AST-2307394 and PHY-2309210, the NSF Frontera supercomputer under
grant AST21006, and Delta at the National Center for Supercomputing Applications
(NCSA) through allocation PHY210074 from the Advanced Cyberinfrastructure
Coordination Ecosystem: Services \& Support (ACCESS) program, which is supported
by National Science Foundation grants \#2138259, \#2138286, \#2138307,
\#2137603, and \#2138296. ERM further acknowledges support on Perlmutter through
NERSC under grant m4575. The authors acknowledge partial support by the Sherman
Fairchild Foundation, and by NSF Grants PHY-2207342 and OAC-2209655 at Cornell.
Figures in this article were produced using Matplotlib \citep{matplotlib} and
Numpy \citep{numpy} packages.

\appendix
\section{Initial conditions for one-dimensional problems}
\label{app:1d tests ID}

\subsection{Fast wave}
Initial conditions are \cite{Komissarov2002}
\begin{equation} \label{eq:fast wave ID}
\begin{split}
    B^x & = 1.0 ,\\
    B^y & = \left\{\begin{array}{ll}
        1.0        & \text{ if } x < -0.1   \\
        -1.5x+0.85 & \text{ if } -0.1 < x < 0.1 \\
        0.7        & \text{ if } x > 0.1
        \end{array}\right\} ,\\
    B^z & = 0 ,\\
    E^x & = 0 ,\\
    E^y & = 0 ,\\
    E^z & = -B^y .
\end{split}
\end{equation}

\subsection{Stationary Alfv\'en wave}
In the rest frame of the wave, electric and magnetic fields are
\cite{Komissarov2004}
\begin{equation} \label{eq:alfven wave initial data}
\begin{split}
    B_x &= 1.0 ,\\
    B_y &= 1.0 ,\\
    B_z & = \left\{\begin{array}{ll}
        1.0         & \text{if } x < -0.1   \\
        1.15 + 0.15 \sin (5\pi x) & \text{if } |x|<0.1 \\
        1.3         & \text{if } x > 0.1
        \end{array}\right\} ,\\
    E_x & = -B_z ,\\
    E_y & = 0 , \\
    E_z & = 1.0 . \\
\end{split}
\end{equation}
The case with nonzero wave speed $-1 < \mu < 1$ can be tested by performing an
appropriate Lorentz boost to the initial conditions \eqref{eq:alfven wave
initial data} \cite[see e.g.][]{Paschalidis2013}.

\subsection{FFE breakdown}

The initial state is \cite{Komissarov2002}
\begin{equation}
\begin{split}
    B^x & = 1 , \\
    B^y & = B^z = \left\{\begin{array}{ll}
        1         & \text{if } x < -0.1   \\
        -10x & \text{if } -0.1 < x < 0.1 \\
        -1         & \text{if } x > 0.1
        \end{array}\right\} , \\
    E^x & = 0 , \\
    E^y & = 0.5 , \\
    E^z & = -0.5 . \\
\end{split}
\end{equation}

\section{Spherical Kerr-Schild coordinates}
\label{sec:spherical kerr-schild coordinates}

The line element of the Kerr spacetime in the Kerr-Schild coordinates is
\begin{equation} \label{eq:kerr metric in ks coordinates}
\begin{split}
ds^2 & = - dt^2 + dx^2 + dy^2 + dz^2
    + \frac{2Mr^3}{r^4 + a^2 z^2}
    \bigg[ \\
    & dt + \frac{r(x dx + y dy) + a(y dx - x dy)}{r^2 + a^2}
        + \frac{z dz}{r} \bigg]^2,
\end{split}
\end{equation}
where $M$ is the mass and $a \equiv J/M^2$ is the dimensionless spin of the
black hole. The coordinate variable $r$ is defined via the relation
\begin{equation}
    \frac{x^2 + y^2}{r^2 + a^2} + \frac{z^2}{r^2} = 1 .
\end{equation}
For $a=0$, we see that $r^2 = x^2 + y^2 + z^2$ is the usual radial coordinate
used in the spherical coordinate system.

In the spherical Kerr-Schild coordinate system, the Kerr metric has the form
\begin{equation}
\label{eq:kerr metric in spherical ks coordinates}
\begin{split}
ds^2 = & -(1-B)\,dt^2 + (1+B)\,dr^2 + \Sigma \, d\theta^2 \\
    & \hspace{1em} + (r^2 + a^2 + B a^2 \sin^2 \theta)\sin^2 \theta \, d\phi^2 \\
    & \hspace{1em} + 2B\,dt dr - 2aB\sin^2 \theta \, dt d\phi \\
    & \hspace{1em} - 2a(1+B) \sin^2 \theta dr d\phi
\end{split}
\end{equation}
where $\Sigma = r^2 + a^2 \cos^2 \theta$ and $B = 2Mr / \Sigma$. In the
spherical Kerr-Schild coordinates, the inner and outer horizon are located at
\begin{equation}
    r_\pm = M(1 \pm \sqrt{1 - a^2}) .
\end{equation}

The coordinate transformation between the Kerr-Schild coordinates \eqref{eq:kerr
metric in ks coordinates} and the spherical Kerr-Schild coordinates
\eqref{eq:kerr metric in spherical ks coordinates} is
\begin{subequations}
\begin{align}
    x & = (r\cos\phi - a \sin\phi) \sin \theta \\
    y & = (r\sin\phi + a \cos\phi) \sin \theta \\
    z & = r \cos \theta
\end{align}
\end{subequations}

\section{The Wald solution}

In the spherical Kerr-Schild coordinates, components of the vector potential
\eqref{eq:exact wald solution} are
\begin{subequations} \label{eq:wald solution components}
\begin{align}
    A_t & = \frac{B_0}{2}(g_{t\phi} + 2a g_{tt}) , \\[.5ex]
    A_r & = \frac{B_0}{2}(g_{r\phi} + 2a g_{tr}) , \\[.5ex]
    A_\theta & = 0                               , \\[.5ex]
    A_\phi & = \frac{B_0}{2}(g_{\phi\phi} + 2a g_{t\phi}).
\end{align}
\end{subequations}
Computing magnetic fields from the vector potential \eqref{eq:wald solution
components}, we get
\begin{subequations}
\begin{align}
\begin{split}
\tilde{B}^r &= B_0 r^2 \sin\theta\cos\theta
    \bigg[ 1  + \frac{a^2}{r^2} \\
    & \hspace{11ex} + \frac{2M}{r}
        \left( \frac{r^4-a^4}{(r^2+a^2\cos^2\theta)^2}-1 \right)
    \bigg]
\end{split}\\[2ex]
\begin{split}
\tilde{B}^\theta &= - B_0 r \sin^2 \theta
    \\ & - \frac{a^2 M B_0 \sin^2 \theta}{(r^2+a^2\cos^2\theta)^2}
        (r^2-a^2\cos^2\theta)(2-\sin^2\theta)
\end{split}\\[2ex]
\begin{split}
\tilde{B}^\phi &= a B_0\sin\theta\cos\theta \left[
    1 + \frac{2Mr (r^2-a^2)}{(r^2+a^2\cos^2\theta)^2}
    \right].
\end{split}
\end{align}
\end{subequations}

We also write out $\tilde{B}^i$ in the Cartesian representation
\begin{subequations}
\begin{align}
    \bar{x} &= r \sin\theta \cos\phi , \\
    \bar{y} &= r \sin\theta \sin\phi , \\
    \bar{z} &= r \cos\theta ,
\end{align}
\end{subequations}
which we use for representing tensor quantities in the code. Resulting
expressions are:

\begin{widetext}
\begin{subequations}
    \label{eq:wald problem-cartesian projection in spherical ks}
\begin{align}
\tilde{B}^{\bar{x}} &= a B_0 \bar{z} \left[
    (a\bar{x}-r\bar{y}) \left\{
        \frac{1}{r^4} + \frac{2M r (r^2-a^2)}{(r^4+a^2z^2)^2}
    \right\}
    + a M r \bar{x} \left\{
        \frac{r^2-z^2}{r^4(r^4+a^2z^2)}
        - \frac{4(r^2+z^2)}{(r^4+a^2z^2)^2}
    \right\}
    \right] \\[1ex]
\tilde{B}^{\bar{y}} &= a B_0 \bar{z} \left[
    (r\bar{x}+a\bar{y}) \left\{
        \frac{1}{r^4} + \frac{2M r (r^2-a^2)}{(r^4+a^2z^2)^2}
    \right\}
    + a M r \bar{y} \left\{
        \frac{r^2-z^2}{r^4(r^4+a^2z^2)}
        - \frac{4(r^2+z^2)}{(r^4+a^2z^2)^2}
    \right\}
    \right] \\[1ex]
\tilde{B}^{z} &= B_0 \left[
    1 + \frac{a^2z^2}{r^4} + \frac{M a^2}{r^3}\left\{
        1 - \frac{z^2(a^2+z^2)(5r^4+a^2z^2)}{(r^4+a^2z^2)^2}
    \right\}
    \right] .
\end{align}
\end{subequations}
\end{widetext}
from which one can check that $\tilde{B}^i \rightarrow (0, 0, B_0)$ for
$a\rightarrow 0$. We use the expressions \eqref{eq:wald problem-cartesian
projection in spherical ks} for initializing the densitized magnetic fields
$\tilde{B}^i$ in the code.

Note that the barred coordinates $\bar{x}, \bar{y}$ are not equal to the $x, y$
coordinates appearing in the original Kerr-Schild form \eqref{eq:kerr metric in
ks coordinates}, whereas $\bar{z} = z$. Barred coordinates $\bar{x}, \bar{y},
\bar{z}$ are simply Cartesian projections of the spherical Kerr-Schild
coordinates \eqref{eq:kerr metric in spherical ks coordinates}, where they are
related with the Kerr-Schild coordinates by
\begin{subequations}
\begin{align}
    \frac{\bar{x}}{r} &= \frac{x}{\sqrt{r^2+a^2}}, \\
    \frac{\bar{y}}{r} &= \frac{y}{\sqrt{r^2+a^2}}, \\
    \bar{z} &= z, \\
    r^2 &= \bar{x}^2 + \bar{y}^2 + \bar{z}^2 .
\end{align}
\end{subequations}

\bibliography{references}
\end{document}